\documentclass[preprint,final,5p,times,twocolumn]{elsarticle}
\usepackage{rotating,color,subfigure,amssymb}
\usepackage{alphalph}
\usepackage{amsmath}
\usepackage[T1]{fontenc}

\usepackage{amssymb}

\journal{Physics Letters B}
\begin{document}

\begin{frontmatter}



\title{Isospin Decomposition of the Basic Double-Pionic Fusion in
  the Region of the ABC Effect}

\author[IKPUU]{The WASA-at-COSY Collaboration\\[2ex] P.~Adlarson}
\author[ASWarsN]{W.~Augustyniak}
\author[IPJ]{W.~Bardan}
\author[PITue]{M.~Bashkanov}
\author[IPJ]{T.~Bednarski}
\author[MS]{F.S.~Bergmann}
\author[ASWarsH]{M.~Ber{\l}owski}
\author[IITB]{H.~Bhatt}
\author[IKPJ,JCHP]{M.~B\"uscher}
\author[IKPUU]{H.~Cal\'{e}n}
% ...

\address[IKPUU]{Division of Nuclear Physics, Department of Physics and 
 Astronomy, Uppsala University, Box 516, 75120 Uppsala, Sweden}
\address[ASWarsN]{Department of Nuclear Physics, National Centre for Nuclear 
 Research, ul.\ Hoza~69, 00-681, Warsaw, Poland}
\address[IPJ]{Institute of Physics, Jagiellonian University, ul.\ Reymonta~4, 
 30-059 Krak\'{o}w, Poland}
\address[PITue]{Physikalisches Institut, Eberhard--Karls--Universit\"at 
 T\"ubingen, Auf der Morgenstelle~14, 72076 T\"ubingen, Germany}
\address[MS]{Institut f\"ur Kernphysik, Westf\"alische Wilhelms--Universit\"at 
 M\"unster, Wilhelm--Klemm--Str.~9, 48149 M\"unster, Germany}
\address[ASWarsH]{High Energy Physics Department, National Centre for Nuclear 
 Research, ul.\ Hoza~69, 00-681, Warsaw, Poland}
\address[IITB]{Department of Physics, Indian Institute of Technology Bombay, 
 Powai, Mumbai--400076, Maharashtra, India}
\address[IKPJ]{Institut f\"ur Kernphysik, Forschungszentrum J\"ulich, 52425 
 J\"ulich, Germany}
\address[JCHP]{J\"ulich Center for Hadron Physics, Forschungszentrum J\"ulich, 
 52425 J\"ulich, Germany}
% ...

\fntext[fnsu]{present address: Department of Physics, Stockholm University, 
 Roslagstullsbacken~21, AlbaNova, 10691 Stockholm, Sweden}
\fntext[fnmz]{present address: Institut f\"ur Kernphysik, Johannes 
 Gutenberg--Universit\"at Mainz, Johann--Joachim--Becher Weg~45, 55128 Mainz, 
 Germany}
\fntext[fnuc]{present address: Department of Physics and Astronomy, University 
 of California, Los Angeles, California--90045, U.S.A.}
\fntext[fnbe]{present address: Albert Einstein Center for Fundamental Physics,
 Fachbereich Physik und Astronomie, Universit\"at Bern, Sidlerstr.~5, 
 3012 Bern, Switzerland}
\fntext[fnsh]{present address: Department of Physics and Astronomy, University 
 of Sheffield, Hounsfield Road, Sheffield, S3 7RH, United Kingdom}

\cortext[coau]{Corresponding author }

\begin{abstract}
Exclusive and kinematically complete high-statistics
measurements of the basic double pionic fusion reactions $pn \to d\pi^0\pi^0$,
$pn~\to~d\pi^+\pi^-$ and $pp \to d\pi^+\pi^0$ have been carried out
simultaneously over the
energy region of the ABC effect using the WASA detector setup at COSY. Whereas
the isoscalar reaction part given by the $d\pi^0\pi^0$ channel exhibits the
ABC effect, {\it i.e.} a low-mass enhancement in the $\pi\pi$-invariant mass
distribution, as well as the associated resonance structure in the total cross
section, the isovector part given by the $d\pi^+\pi^0$ channel shows a smooth
behavior consistent with the conventional $t$-channel $\Delta\Delta$
process. The $d\pi^+\pi^-$ data are very well reproduced by combining the data
for isovector and isoscalar contributions, if the kinematical consequences of
the isospin violation due to different masses for charged and neutral pions
are taken into account. 
\end{abstract}


\end{frontmatter}





\section{Introduction}
The so-called ABC effect denotes a pronounced low-mass enhancement in
the $\pi\pi$-invariant mass spectrum of double-pionic fusion reactions. It is
named after Abashian, Booth and Crowe \cite{abc}, who first observed it in the
inclusive measurement of the $pd \to ^3$He X reaction in the kinematic region 
corresponding to the production of two pions. Recent exclusive measurements of
the $pn \to d\pi^0\pi^0$ reaction revealed the ABC effect to be associated
with a narrow resonance structure in the energy dependence of the 
total $pn \to d\pi^0\pi^0$ cross section \cite{prl2011,MB}. For this resonance
structure the quantum numbers $I(J^P) = 0(3^+)$ have been determined as well
as a mass of m = 2.37 GeV. The latter is about 80 MeV below the nominal mass of
$2m_{\Delta}$ of a conventional mutual excitation of the two participating
nucleons into their $\Delta(1232) P_{33}$ state by $t$-channel meson exchange
\cite{ris}. The observed width of only about 70 MeV is more than three times
smaller than that of the conventional $\Delta\Delta$ process.  

In this scenario the isovector
double-pionic fusion reaction $pp \to d \pi^+\pi^0$ should exhibit neither an
ABC effect in the $\pi^+\pi^0$-invariant mass spectrum nor an ABC resonance
structure in the total cross section. So far there has been only one 
exclusive and kinematically complete measurement of this reaction performed at
CELSIUS at a beam energy of 1.1 GeV \cite{FK}. The measured
$\pi^+\pi^0$-invariant mass spectrum does not show any low-mass enhancement,
but rather a low-mass suppression. The latter is in accordance with the
constraint from Bose symmetry that an isovector pion pair cannot be in
relative $s$-wave, but in relative $p$-wave. In addition to this measurement there are few
low-statistics bubble chamber total cross section data spread over the energy
region from threshold up to $\sqrt s$ = 3 GeV \cite{shim,bys}. All these data
are consistent with a $t$-channel $\Delta\Delta$ process
 \cite{FK}. This process has been shown to be the leading two-pion
production process at beam energies above 1 GeV \cite{alv,iso,deldel,nnpipi}
- at least for $pp$ induced two-pion production.

The third basic double-pionic fusion reaction, the $pn \to d \pi^+\pi^-$
reaction, contains both isoscalar and isovector contributions. From isospin
conservation we expect for the angle integrated cross sections \cite{bys}

\begin{equation}
\sigma(pn \to d \pi^+\pi^-) = 2 \sigma(pn \to d\pi^0\pi^0) + \frac {1}{2}
\sigma(pp \to d\pi^+\pi^0). 
\end{equation}

For this reaction there are bubble chamber measurements from DESY \cite{bar}
and JINR Dubna \cite{abd},
which also provide some low-statistics differential distributions. However,
their statistical precision is not good enough to clearly decide whether the
ABC effect is  present in this reaction or not.  On the other hand inclusive
single-arm magnetic 
spectrometer measurements of high statistics show convincing evidence for the
presence of the ABC effect in this reaction \cite{plo}.

\section{Experiment}

We have carried out exclusive and kinematically complete
measurements of these three basic double-pionic fusion reactions over the main
region of the ABC effect by impinging a proton beam of $T_p$ = 1.2
GeV on the
deuterium pellet target of the WASA detector facility at COSY
\cite{barg,wasa}. By the simultaneous observation of all three reaction channels
systematic uncertainties in the measurements are minimized.
The quasi-free reactions $pd \to d \pi^0\pi^0 +
p_{spectator}$, $pd \to d \pi^+\pi^- + p_{spectator}$ and $pd \to d \pi^+\pi^o
+ n_{spectator}$ have been selected by requiring a deuteron track in the
forward detector as well as signals induced by pions in the central detector. The tracks
from charged pions, which are bent due to the magnetic field supplied by a
superconducting solenoid, have been reconstructed from the hit patterns
recorded by a mini drift chamber consisting of 1738 drift tubes (straws). The
photons originating from the $\pi^0$ decay have been registered in a
scintillating electromagnetic calorimeter consisting of 1012 sodium doped CsI
crystals. 

That way the four-momentum of the unobserved spectator nucleon has been
reconstructed by applying a kinematic fit with three, one and 
two overconstraints for events originating from $pd \to d \pi^0\pi^0 +
p_{spectator}$, $pd \to d \pi^+\pi^- + p_{spectator}$ and $pd \to d \pi^+\pi^o
+ n_{spectator}$ reactions, respectively. The measured spectator momentum
distributions are as shown in Fig.~1 of Ref. \cite{prl2011} for the case of
the  $pd \to d \pi^0\pi^0 + p_{spectator}$ reaction. As in
Ref. \cite{prl2011} we only use spectator momenta $<$ 0.16 GeV/c for the
further data analysis. This implies 
an energy range of 2.3 GeV $< \sqrt s <$ 2.5 GeV being covered due to the
Fermi motion of the nucleons in the target deuteron. For a target at rest this
energy range would correspond to incident lab energies of 1.07 GeV $< T_p < $
1.39 GeV. 


\begin{figure} 
\centering
\includegraphics[width=1\columnwidth]{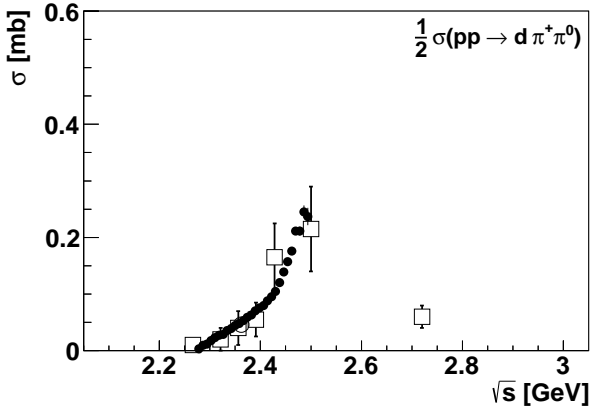}
\includegraphics[width=1\columnwidth]{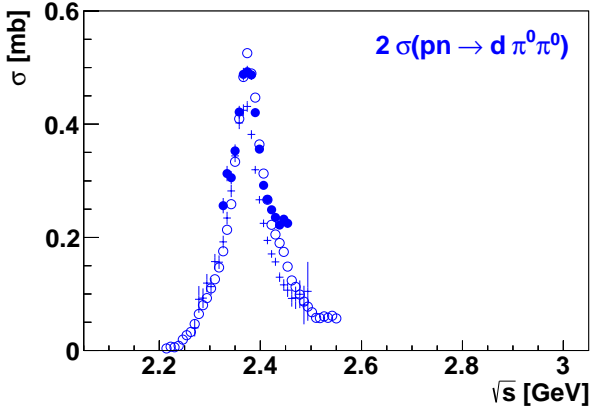}
\includegraphics[width=1\columnwidth]{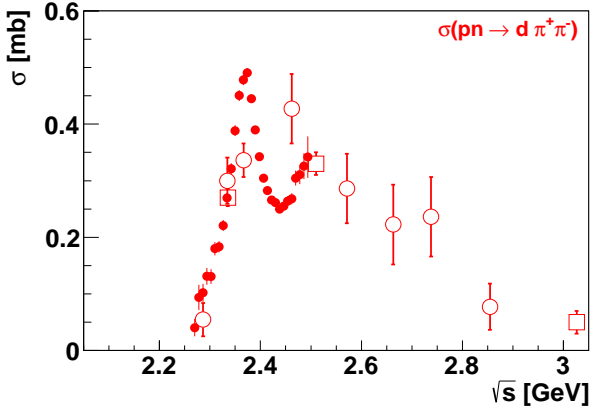}
\caption{
Total cross sections of the basic double-pionic fusion reactions $pN
  \to d \pi\pi$ of different isospin systems in dependence of the
  center-of-mass energy $\sqrt s$ from threshold ($\sqrt s$ = 2.15 GeV) until
  3.2 GeV. Top: the purely isovector reaction $pp \to d\pi^+\pi^0$, middle:
  the purely isoscalar reaction $pn \to d\pi^0\pi^0$, bottom: the isospin
  mixed reaction  $pn \to d\pi^+\pi^-$. Filled symbols denote results from 
  this work, open symbols from previous work
  \cite{FK,shim,bys,prl2011,bar,abd}. The results from Ref. \cite{prl2011} have
  been renormalized -- see text. The crosses denote the result for the
  $pn \to d\pi^0\pi^0$ reaction by using eq. (1) with the data for the $pn \to
  d\pi^+\pi^-$ and $pp \to d\pi^+\pi^0$ channels as input. 
}
\label{fig1}
\end{figure}

\begin{figure} 
\centering
\includegraphics[width=1\columnwidth]{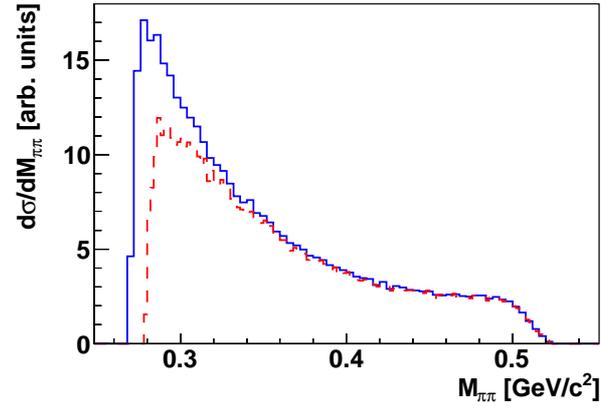}
\caption{Distribution of the simulated $\pi\pi$-invariant mass in the ABC
  region for the 
  production of a $\pi^0\pi^0$ (solid) and a $\pi^+\pi^-$ (dashed) pair,
  respectively. Shown are (acceptance and efficiency corrected) ABC model
  calculations \cite{prl2011} for $\sqrt s$ 
  = 2.37 GeV. The different thresholds at the low-mass side 
  due to the mass difference between charged and neutral pions lead
  to a substantial difference in the spectra and consequently of the
  corresponding cross sections.  
}
\label{fig2}
\end{figure}

Efficiency and acceptance corrections of the data have been determined by MC
simulations of reaction process and detector setup. The absolute normalization
of the data could be done as in previous measurements \cite{prl2011,MB} by
normalizing to the simultaneously measured quasi-free $\eta$ production  $pd
\to d \eta + p_{spectator}$ together with its $3\pi^0$ decay and comparison to
previous results \cite{calen}. However, since the $\eta$ production threshold
is just in the middle of the energy range considered here and its cross
section is still very small in this energy range, this control channel is of
correspondingly low statistics. Taking into account the substantial systematic
uncertainties associated with this threshold region normalization
uncertainties get as large as 30 - 50 $\%$. Therefore we normalize our data in
absolute scale to the Dubna datum for the $np 
\to d\pi^+\pi^-$ reaction at $\sqrt s$ = 2.33 GeV ($T_n$ = 1.03 GeV, open
square symbol in Fig.~1, bottom)
\cite{abd}, which is $\sigma$ = 0.270(15) mb obtained by use of a
neutron beam with a 1$\%$ momentum resolution. 
That way we also 
achieve simultaneously good 
agreement with the second Dubna point at  $\sqrt s$ = 2.51 GeV, -- see Fig.~1,
bottom. We find also good agreement to all previous total
cross section results for the $pp \to d\pi^+\pi^0$ reaction \cite{FK,shim,bys}
-- see Fig.~1, top.
 
Concerning the absolute scale for the $pn \to d\pi^0\pi^0$ reaction there are
seemingly large discrepancies to the previously published data
\cite{prl2011}. In those data there is a 30$\%$ discrepancy in absolute
normalization between data sets taken at $T_p$ = 1.2 and 1.4 GeV. Whereas
the lower energy data cover just the $\eta$ threshold region as discussed above,
the higher energy data are already substantially above this threshold, where
the $\eta$ production cross section is already close to saturation with a much
smaller energy dependence. This situation looks much more
reliable for an absolute normalization and readjusting the data taken at $T_p$
= 1.0 and 1.2 GeV to the 1.4 GeV data lowers the peak cross section already to
0.34 mb. In addition the $\eta$ production data \cite{calen}, where we have
normalized to, have themselves an absolute normalization uncertainty of
additional 30$\%$, so that finally our new result of a peak cross section of
0.27 mb is not at variance with the previous result. The new result means a
renormalization of the previous $T_p$ = 1.4 GeV data by a factor of 0.79 and
of the previous data taken at $T_p$ = 1.0 and 1.2 GeV by a factor of 0.63.

\section{Results and Discussion}

Fig.~1 exhibits the observed energy dependence of the total cross section for
the three double-pionic fusion reactions to the deuteron. The results
of this work are given by the filled symbols and compared to previous
measurements (open symbols) at CELSIUS \cite{FK}, KEK \cite{shim,bys},
COSY\cite{prl2011}, DESY \cite{bar} and JINR Dubna \cite{abd}.

On top of Fig.~1 the purely 
isovector reaction $pp \to d\pi^+\pi^0$ is shown. In the  energy
region covered by this experiment our data exhibit a smooth monotonical
rise of the cross section with energy -- in good agreement with previous
data. Over the full energy region the measurements exhibit a broad structure,
which is taken into account very well by calculations of the
$t$-channel $\Delta\Delta$ process, which produces a resonance-like structure
peaking at $\sqrt s \approx 2 m_{\Delta}$ with a width of about 230 MeV, {\it
  i.e.} twice the $\Delta$ width -- see Fig.~6 in Ref. \cite{FK}.

For the isospin mixed (I = 0 and 1) reaction $pn \to d \pi^+\pi^-$ the
observed energy dependence of the total cross section is 

\onecolumn{
\begin{figure}
\begin{center}
\includegraphics[width=0.49\textwidth]{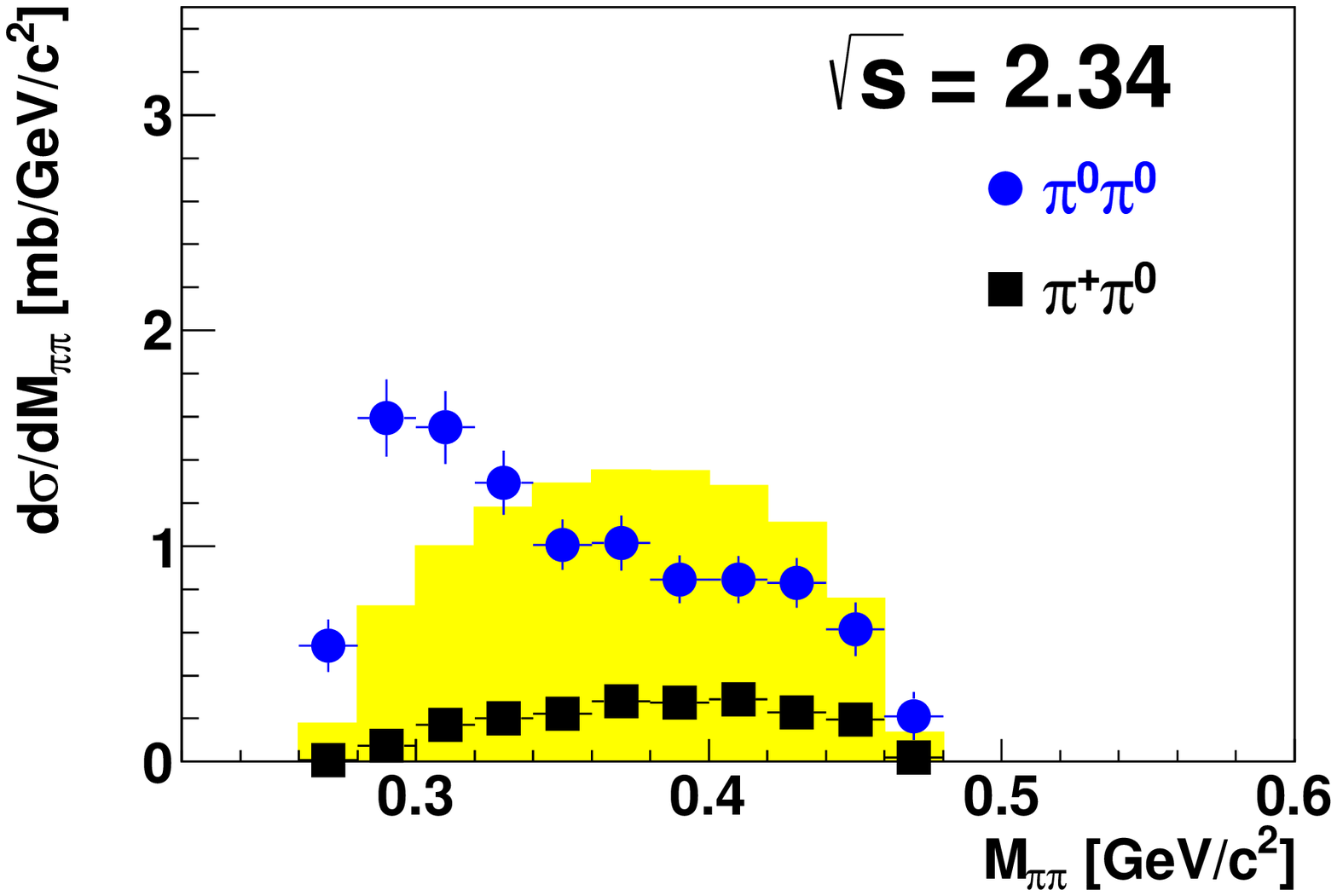}
\includegraphics[width=0.49\textwidth]{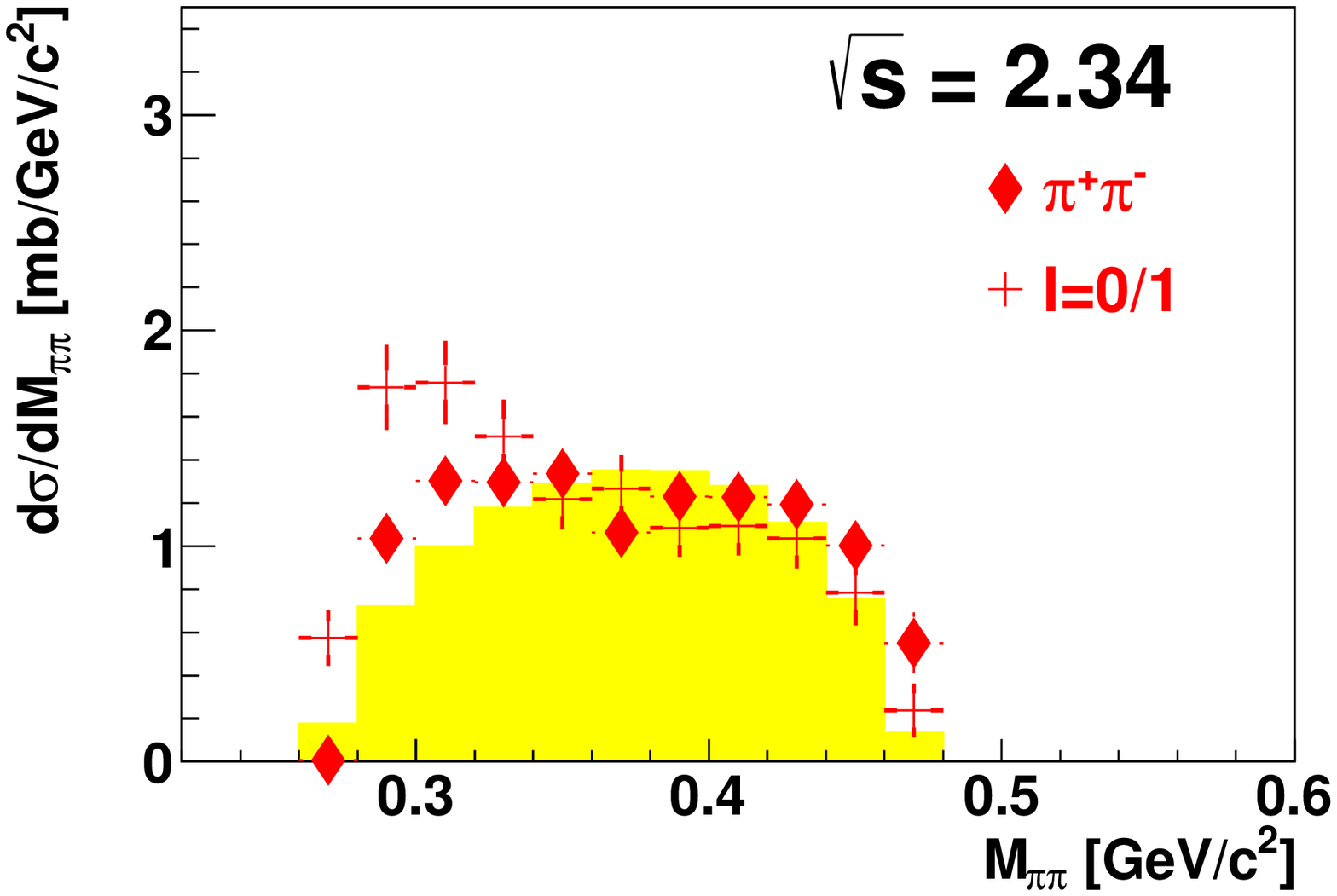}\\
\includegraphics[width=0.49\textwidth]{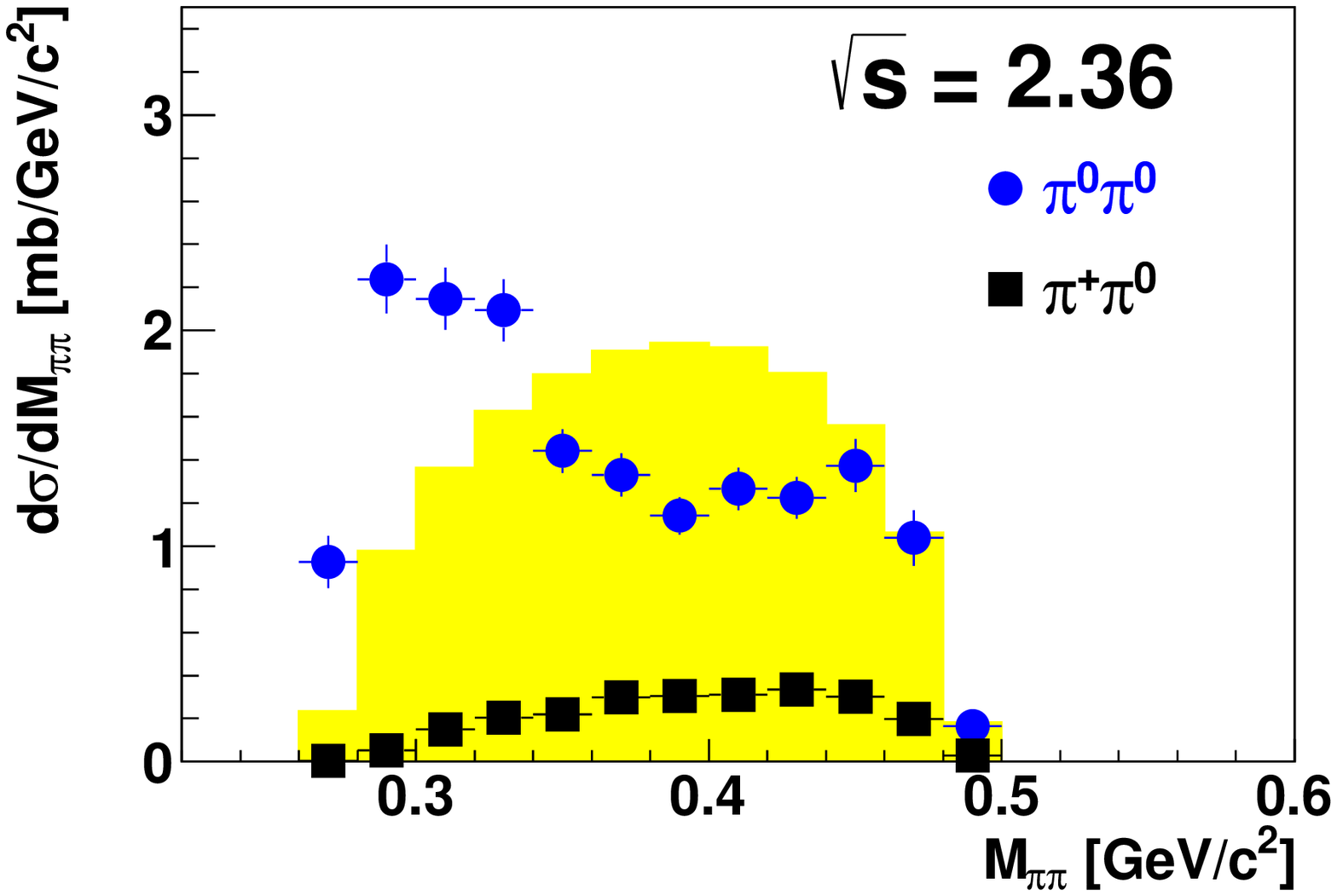}
\includegraphics[width=0.49\textwidth]{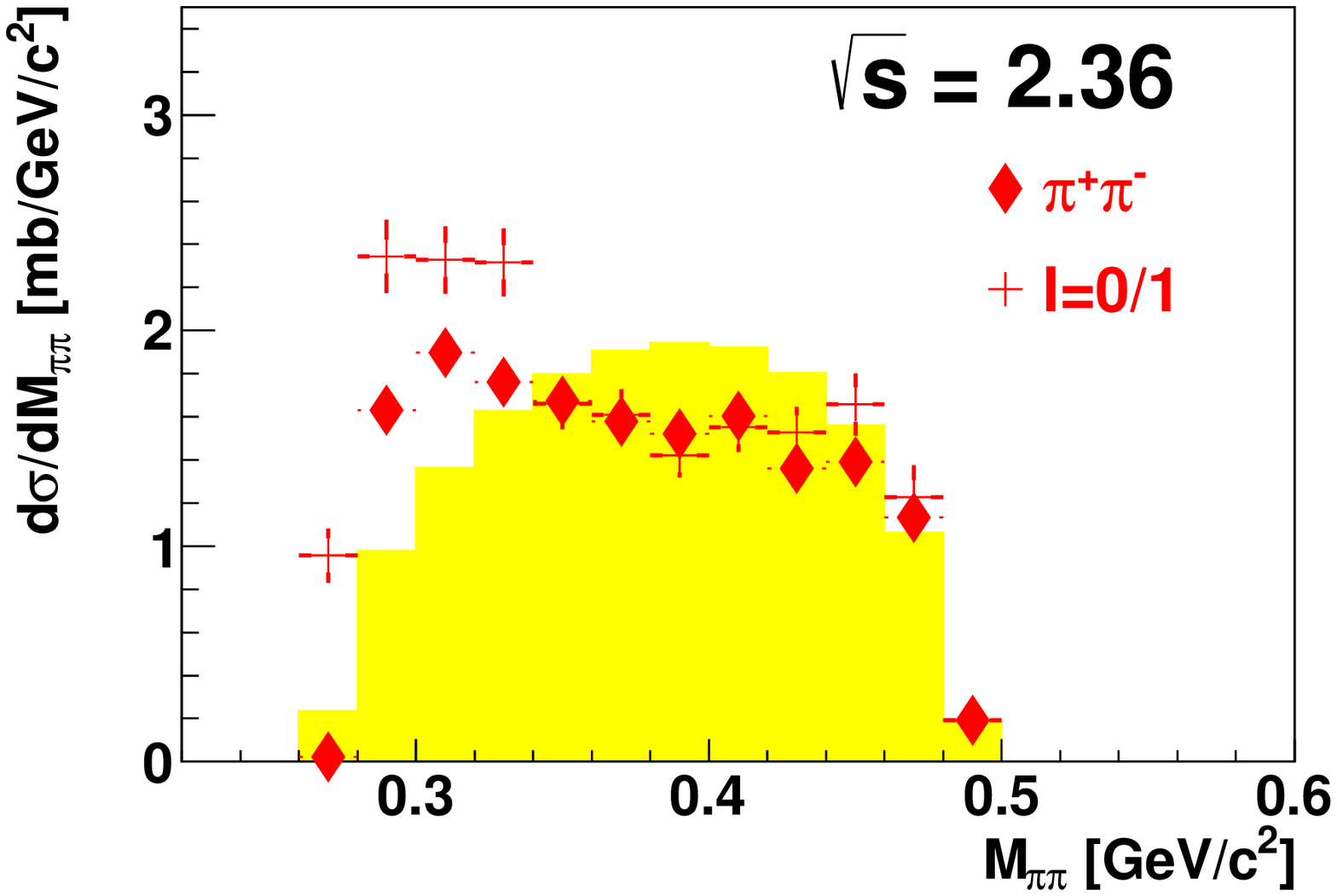}\\
\includegraphics[width=0.49\textwidth]{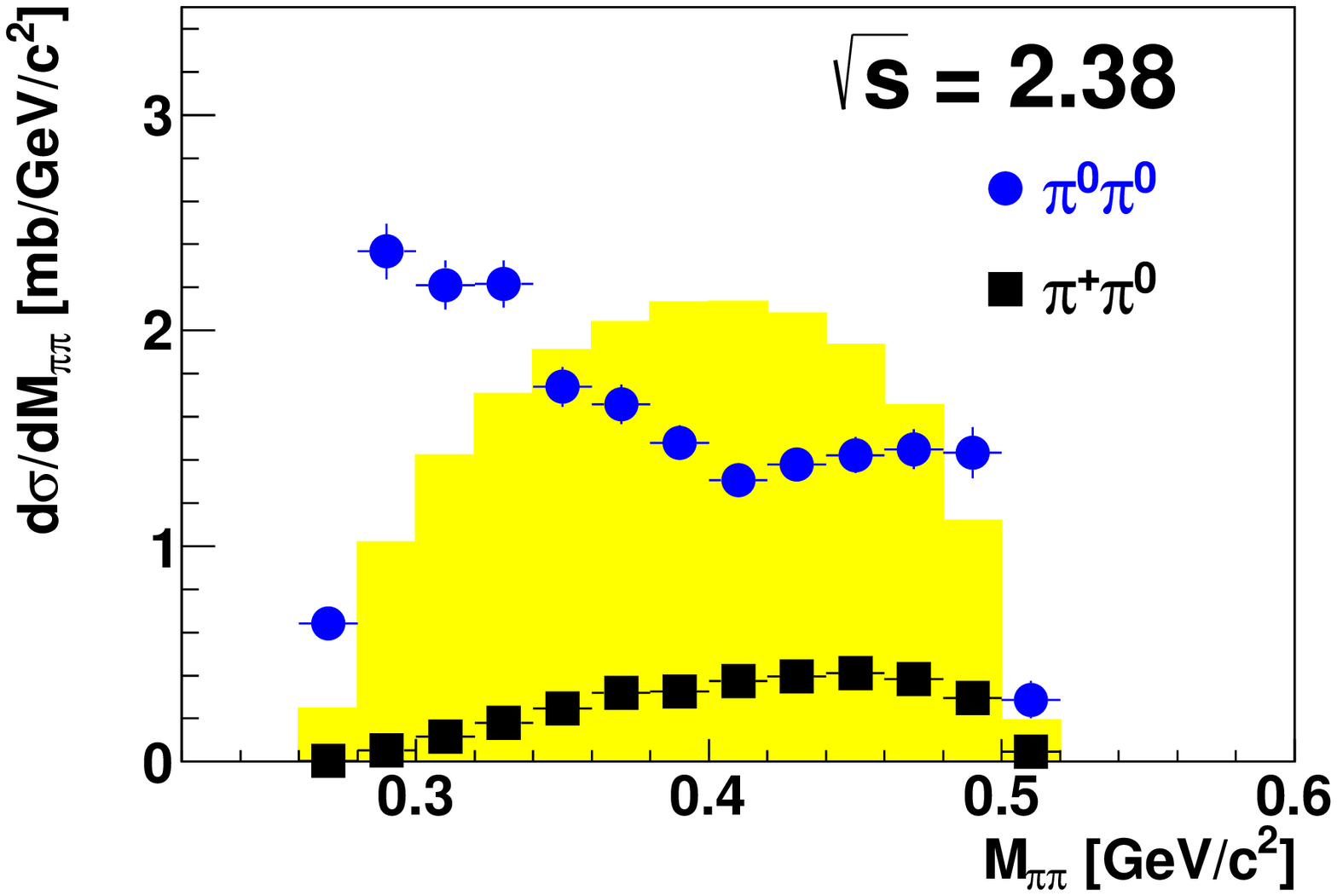}
\includegraphics[width=0.49\textwidth]{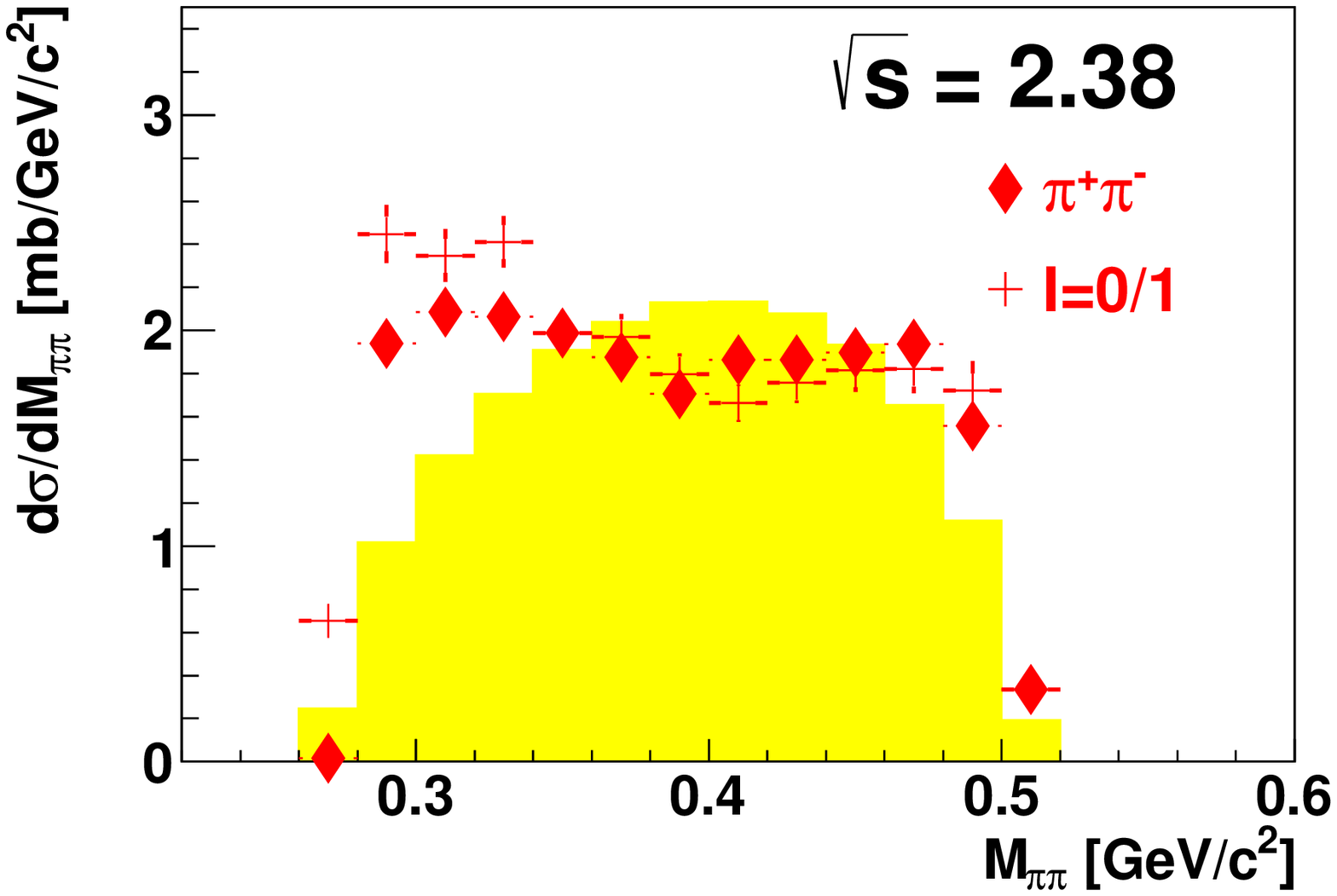}\\

\caption{ 
   Distribution of the $\pi\pi$-invariant mass for the three basic
   double-pionic fusion reactions at $\sqrt s$ = 2.34 GeV, 2.36 and 2.38
   GeV using an energy bin width of 0.02 GeV. On the left the results are shown for the $\pi^0\pi^0$ 
   (circles) and $\pi^+\pi^0$ (squares) systems multiplied by isospin factors
   2 and $\frac{1}{2}$, respectively -- see equation (1). On the right the
   results for the isospin-mixed $\pi^+\pi^-$ system are shown. Data from the
   $pn \to d\pi^+\pi^-$ measurement are given by diamonds, whereas the sum of
   isoscalar and isovector contributions according to equation (1) is shown by
   crosses. The light-shaded areas denote phase space distributions. 
}
\label{fig3}
\end{center}
\end{figure}

\begin{figure}[t]
\begin{center}
\includegraphics[width=0.49\textwidth]{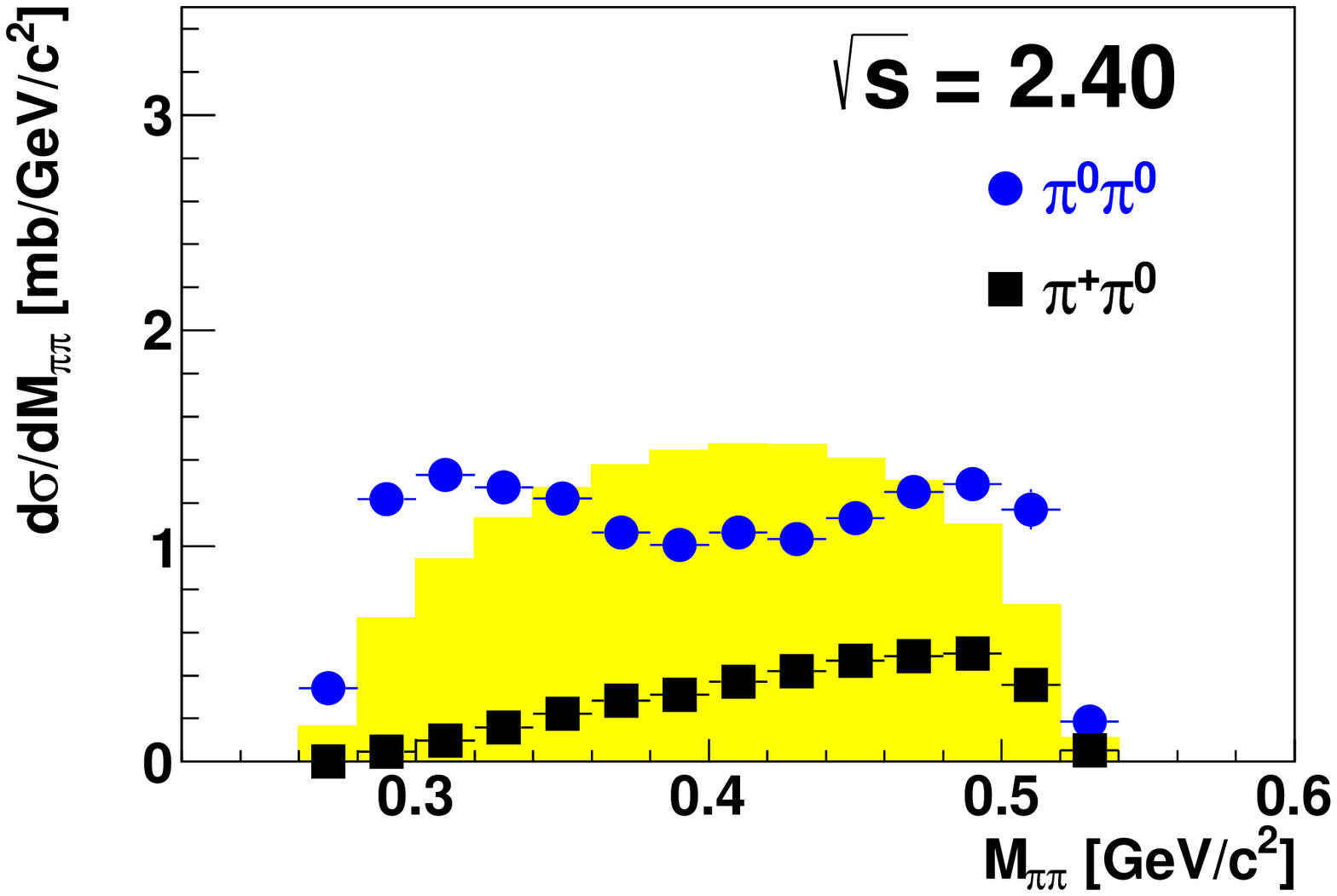}
\includegraphics[width=0.49\textwidth]{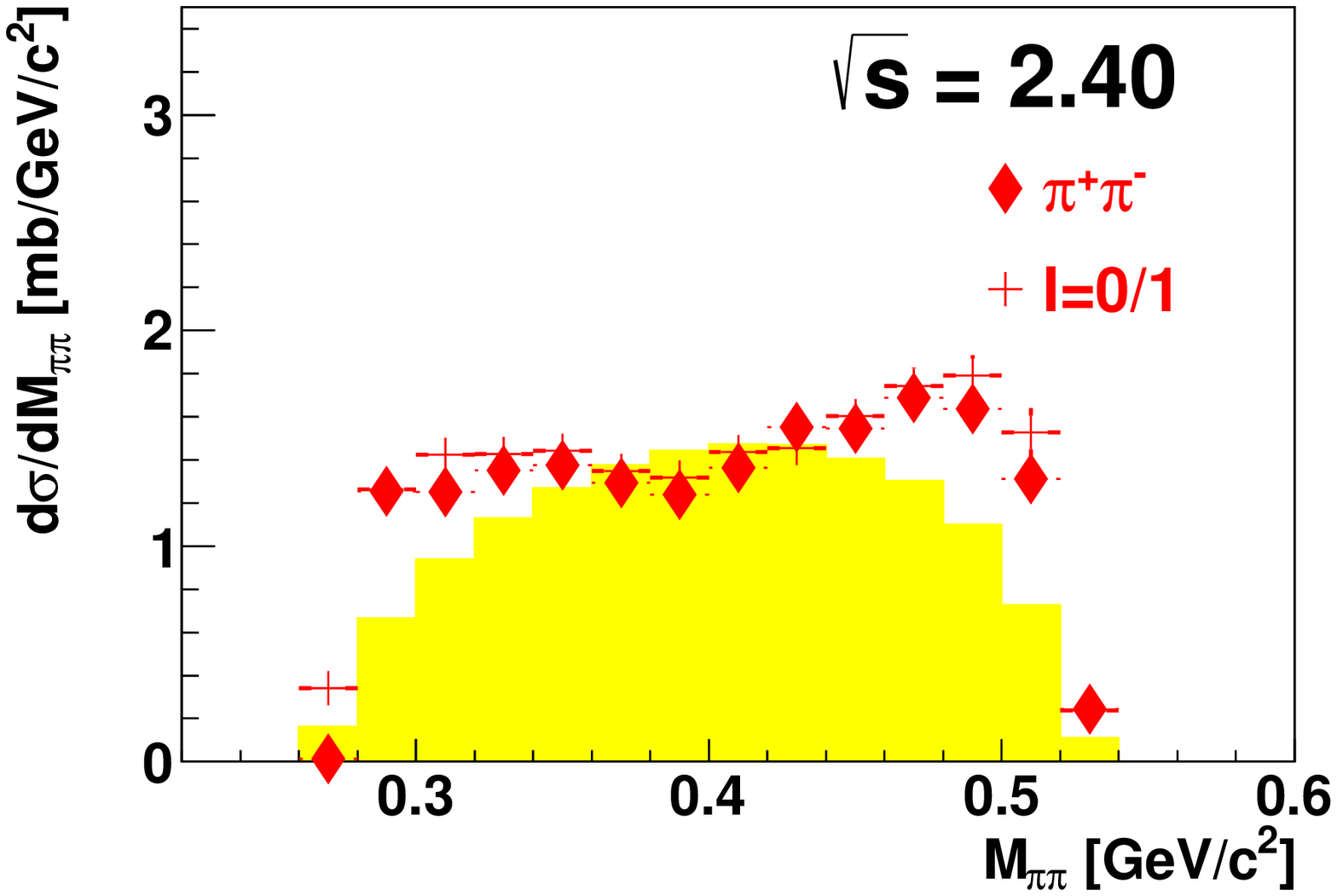}\\
\includegraphics[width=0.49\textwidth]{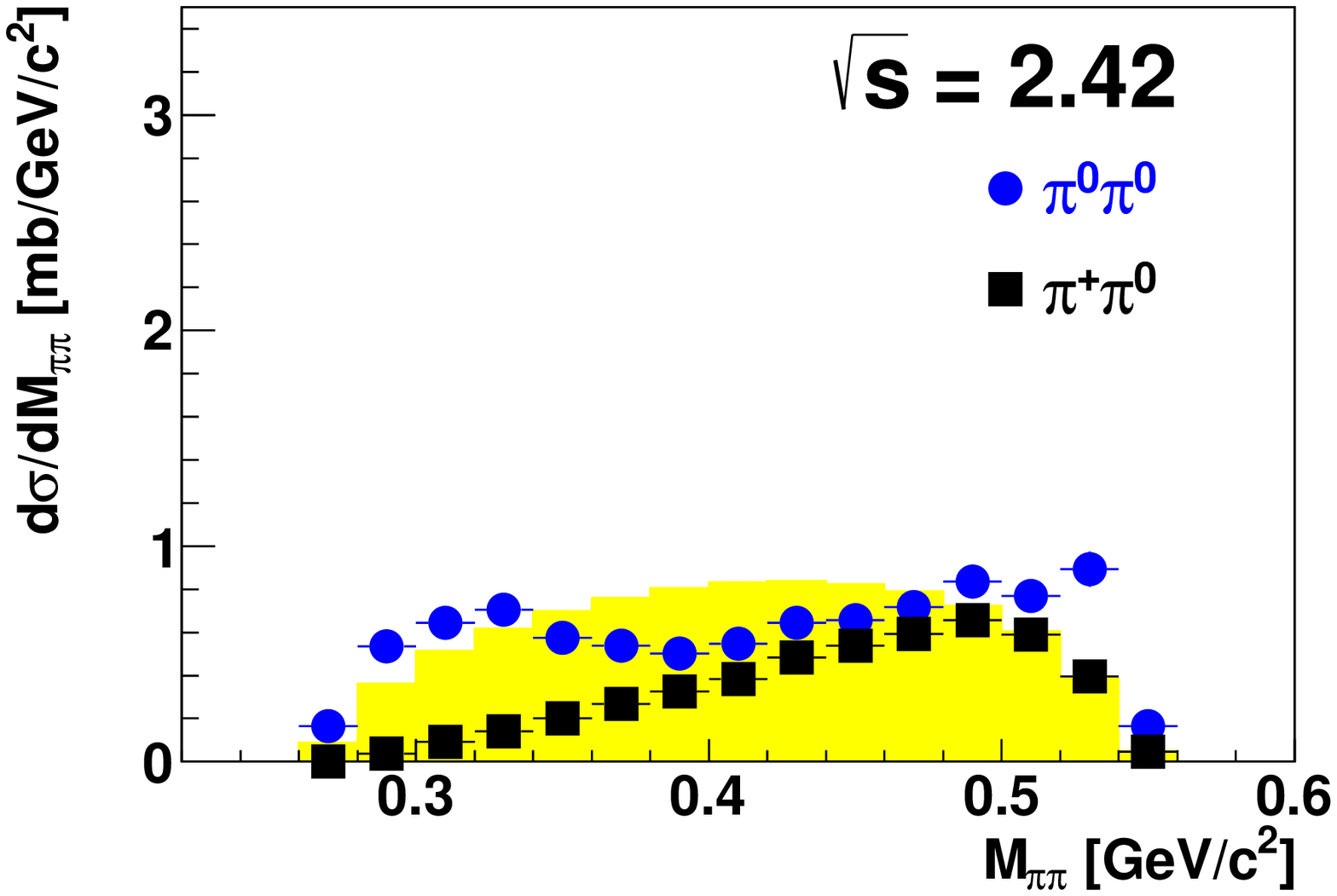}
\includegraphics[width=0.49\textwidth]{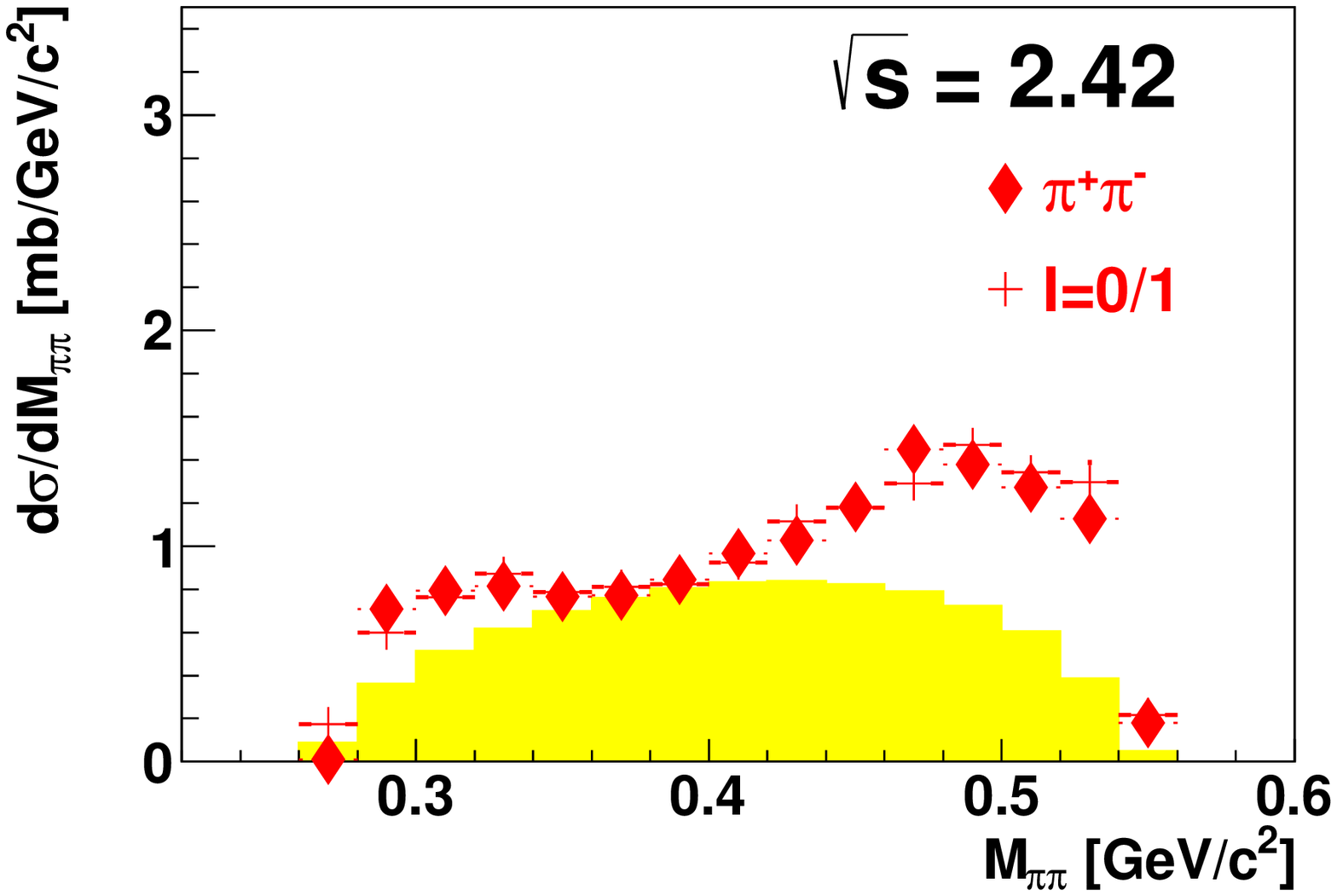}\\
\includegraphics[width=0.49\textwidth]{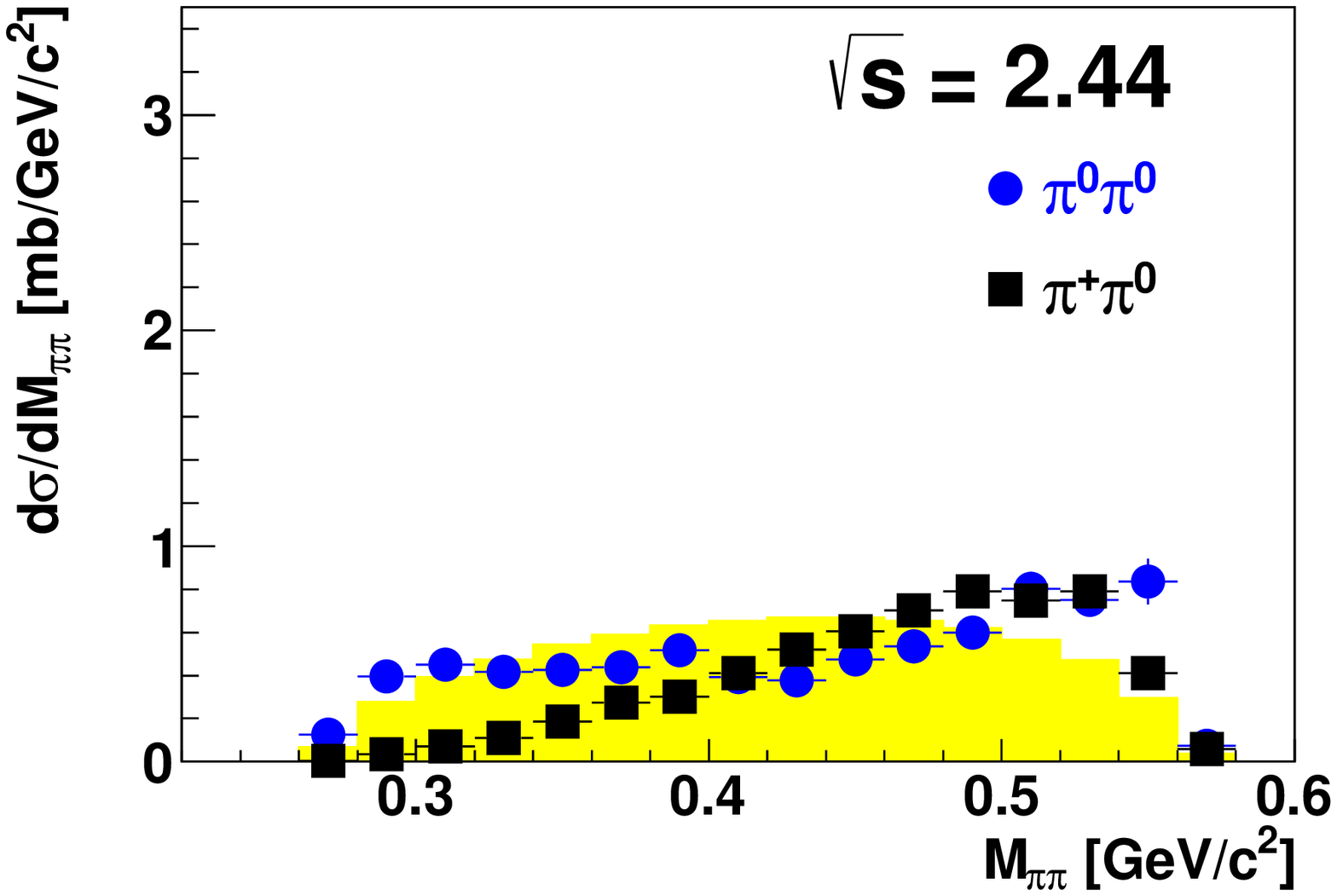}
\includegraphics[width=0.49\textwidth]{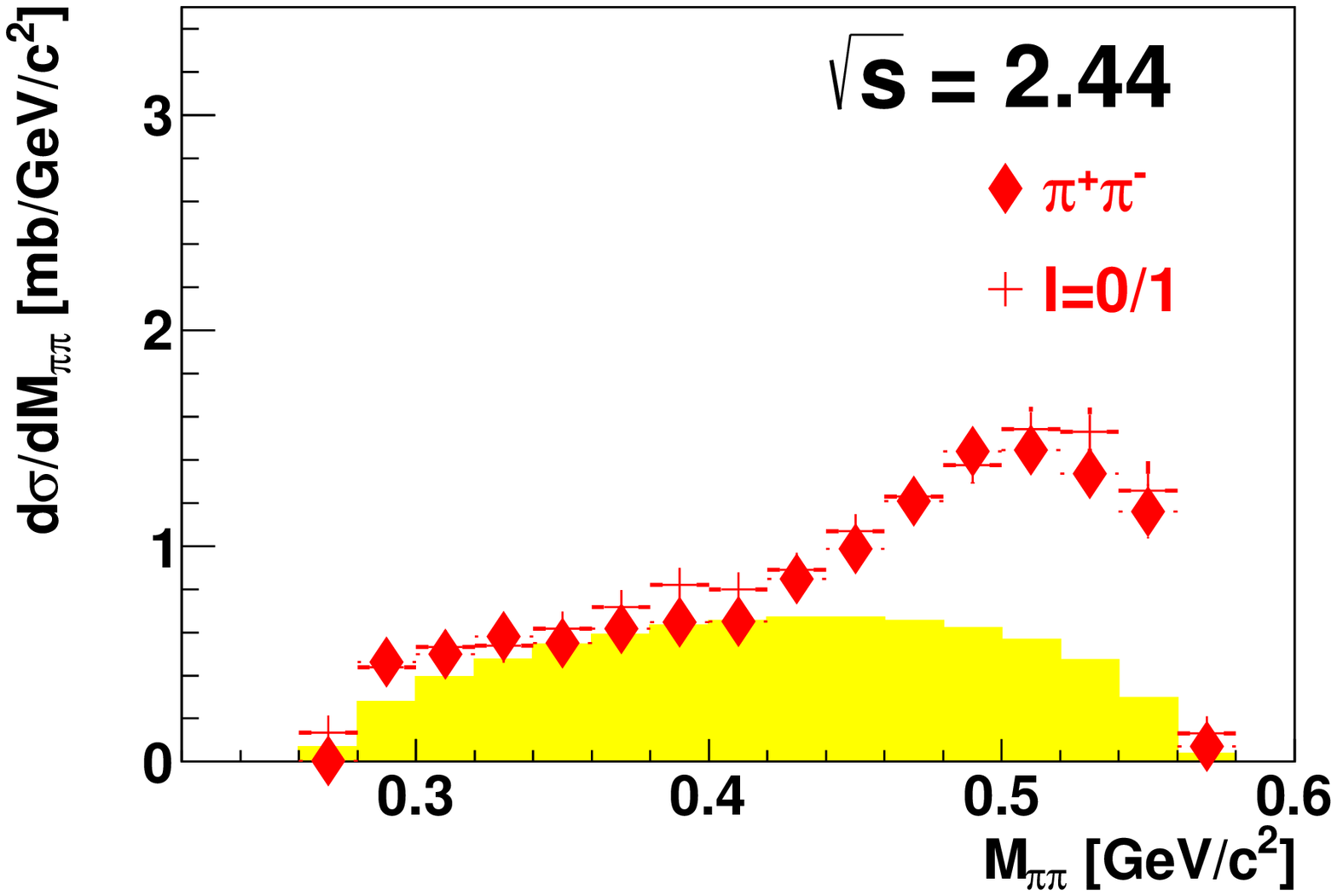}

\caption{The same as Fig.~3, but for $\sqrt s$ = 2.40, 2.42 and 2.44 GeV}
\label{fig3}
\end{center}
\end{figure}
}
\twocolumn{


depicted at the
bottom of Fig.~1. Open symbols refer to previous bubble chamber
measurements at DESY (circles) \cite{bar} and Dubna (squares) \cite{abd}. Our
new data for this channel fit well to the previous results, which have been
obtained partially with a neutron beam of substantial energy spread --
particularly in case of the DESY data. Hence it is not surprising that in the
latter the narrow structure around $\sqrt s$ = 2.37 GeV appearing in our data
is not seen.  

The middle part in Fig.~1 shows the results for the purely
isoscalar case given by the $pn \to d\pi^0\pi^0$ reaction. Open 
symbols denote
our previous measurement \cite{prl2011}. In order to get best overlap with
our present results (filled circles), the previous data taken at $T_p$ = 1.0
and 1.2 GeV have been rescaled by 0.63 
and those taken at $T_p$ = 1.4 GeV by 0.79 --- 
see discussion on absolute normalization in the previous section. Also shown
are the results (crosses) for the extraction of the isoscalar cross section 
from the measured $pn \to d \pi^+\pi^-$ and $pp \to d\pi^+\pi^0$ cross
sections by use of the isospin relation eq. (1). These derived results are
consistent in shape with the directly measured ones. However, they are
systematically smaller with increasing energy. The reason for this is 
the isospin violation in the pion 
mass. Since the mass of two neutral pions is 10 MeV smaller than that of two
charged pions the available phase spaces for neutral and charged pion pairs
differ accordingly. Since in addition the ABC effect and its associated
form-factor in the model description \cite{prl2011} push the strength
distribution in the $M_{\pi\pi}$  spectra towards the low-mass threshold, this
kinematic effect gets substantial, as displayed in Fig.~2, reaching 25$\%$ in
the integrated cross section, {\it i.e.} in total cross section. Hence all the 
three data sets are consistent with each other -- and the narrow resonance
structure, which was determined in Ref.~\cite{prl2011} to have $I(J^P) =
0(3^+)$, appears in both the purely isoscalar $pn \to d \pi^0\pi^0$ and in the
isospin-mixed $pn \to d \pi^+\pi^-$ reaction, but not in the purely isovector
$pp \to d\pi^+\pi^0$ channel.

The measured $M_{\pi\pi}$ distributions are shown in Figs.~3 and 4 for six 
energy bins across the measured energy region. The bin
width here and in all plots shown in subsequent figures is 20 MeV. On the left
side the  $M_{\pi^0\pi^0}$ (circles) and $M_{\pi^+\pi^0}$ (squares)
distributions as derived from 
the $pn \to d\pi^0\pi^0$ and  $pp \to d\pi^+\pi^0$ reactions are shown. They
have been multiplied by the isospin factors 2 and $\frac {1} {2}$,
respectively, in order to give the proper isoscalar and isovector
contributions as needed for the isospin decomposition of the $pn \to
d\pi^+\pi^-$ reaction, see eq.~(1). The observed $M_{\pi^+\pi^-}$ spectra for
the latter reaction are shown by diamonds at the right side in Fig.~3. They
are compared to the sum (crosses) of isoscalar and isovector contributions,
as plotted on the left side. Again we find good agreement between
the directly measured $\pi^+\pi^-$ data and the ones reconstructed from the
two isospin components. The only major difference is at the kinematic
threshold at low masses due to the pion mass effect discussed above and shown
in Fig.~2. 

\begin{figure}
\begin{center}

\includegraphics[width=0.42\textwidth]{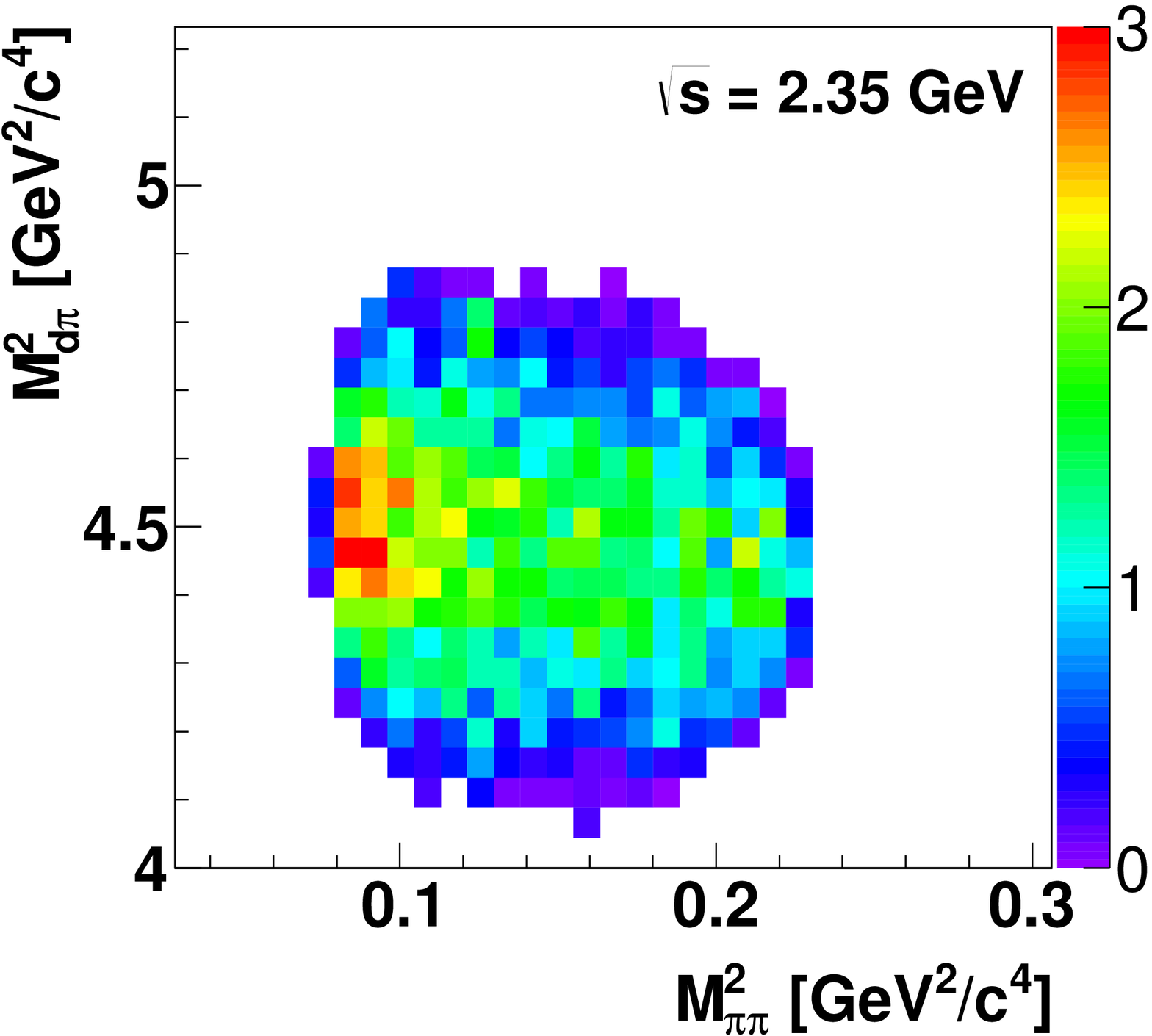}
\includegraphics[width=0.42\textwidth]{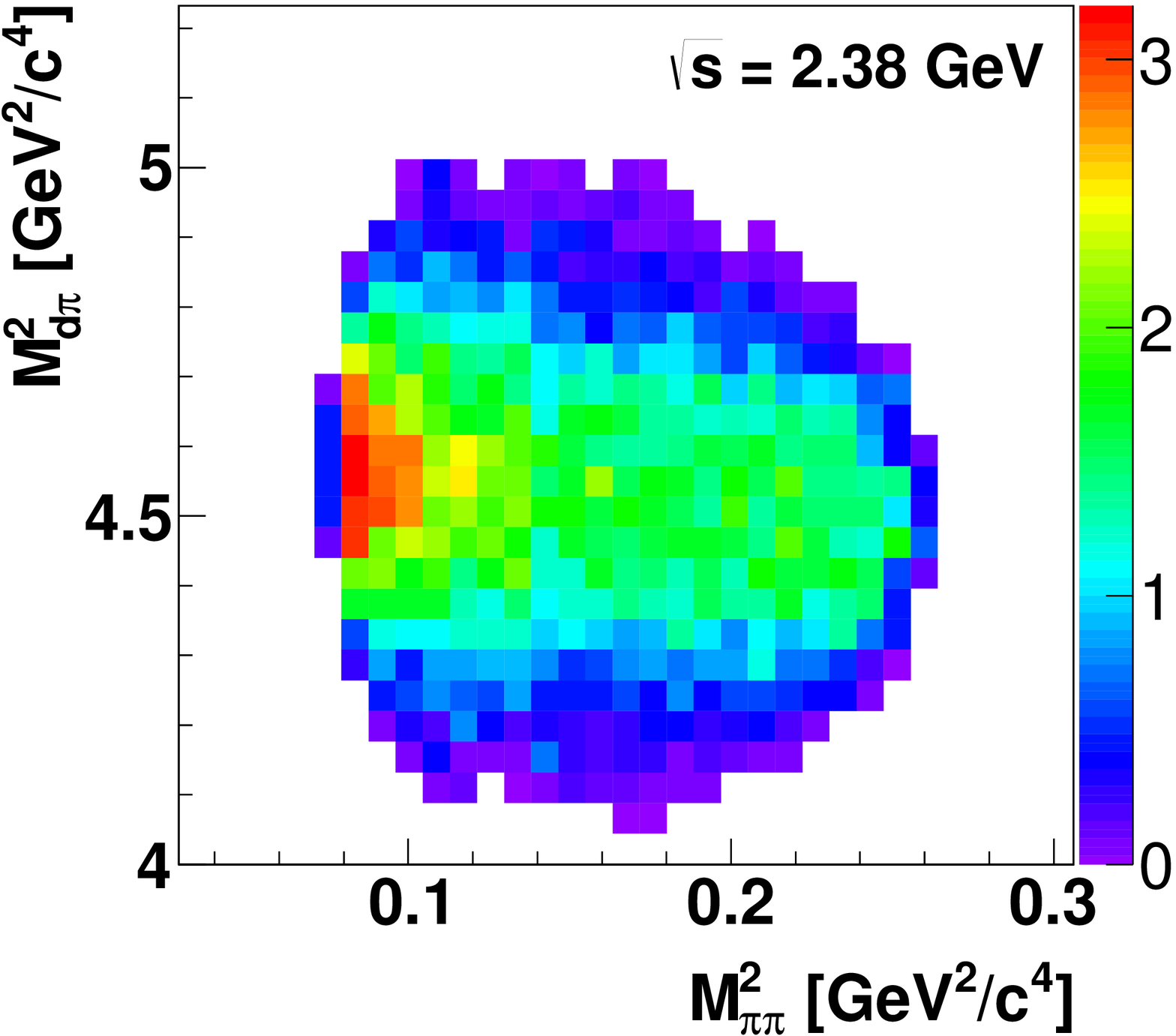}
\includegraphics[width=0.42\textwidth]{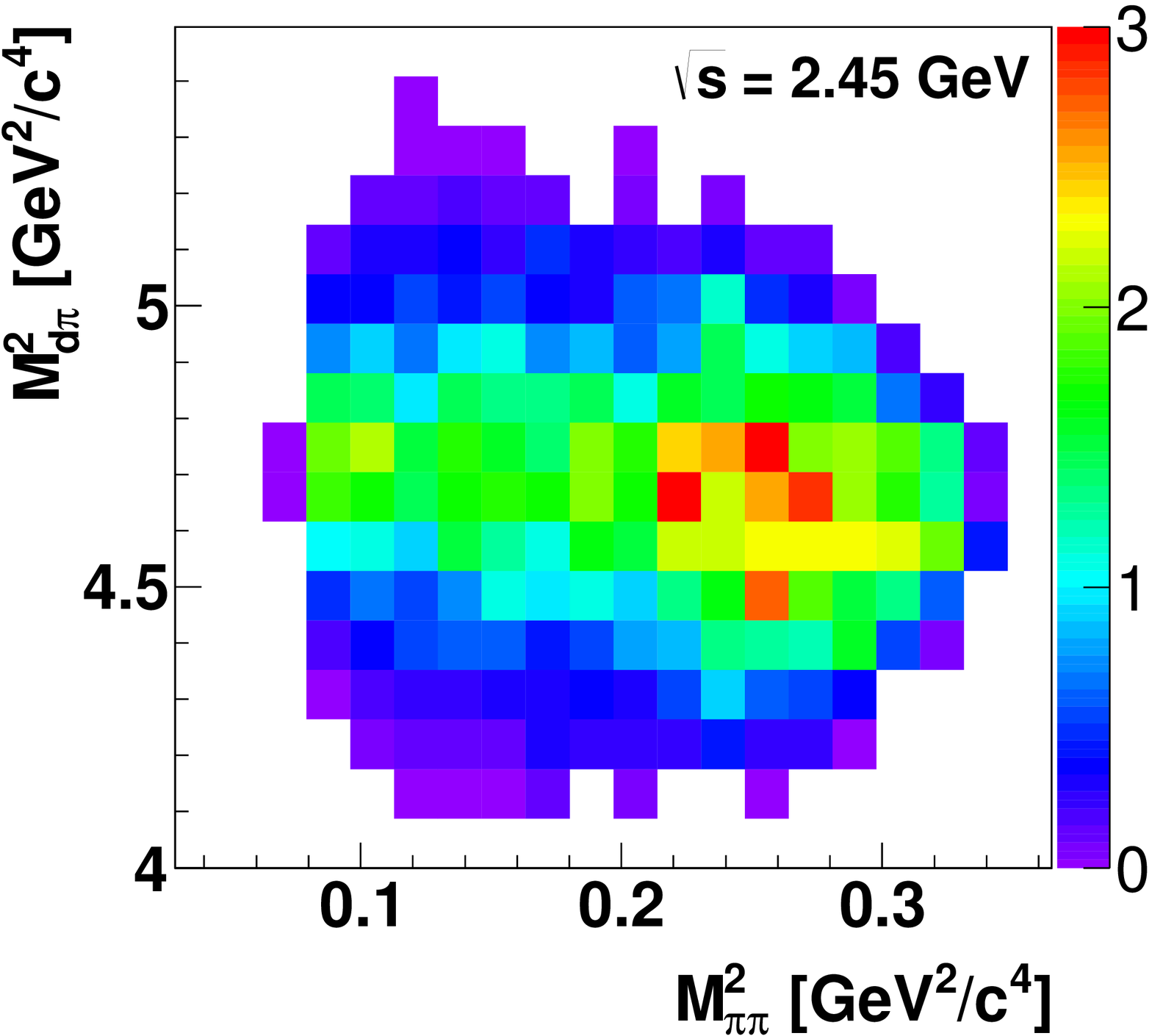}

\caption{
  Dalitz plot of $M_{d\pi^+}^2$ versus $M_{\pi^+\pi^-}^2$ for the $pn \to
  d\pi^+\pi^-$ reaction at $\sqrt s$~=~2.35~GeV (top), 2.38 GeV (middle) and
  2.45 GeV (bottom).
}
\label{fig4}
\end{center}
\end{figure}

\begin{figure}
\begin{center}

\includegraphics[width=0.42\textwidth]{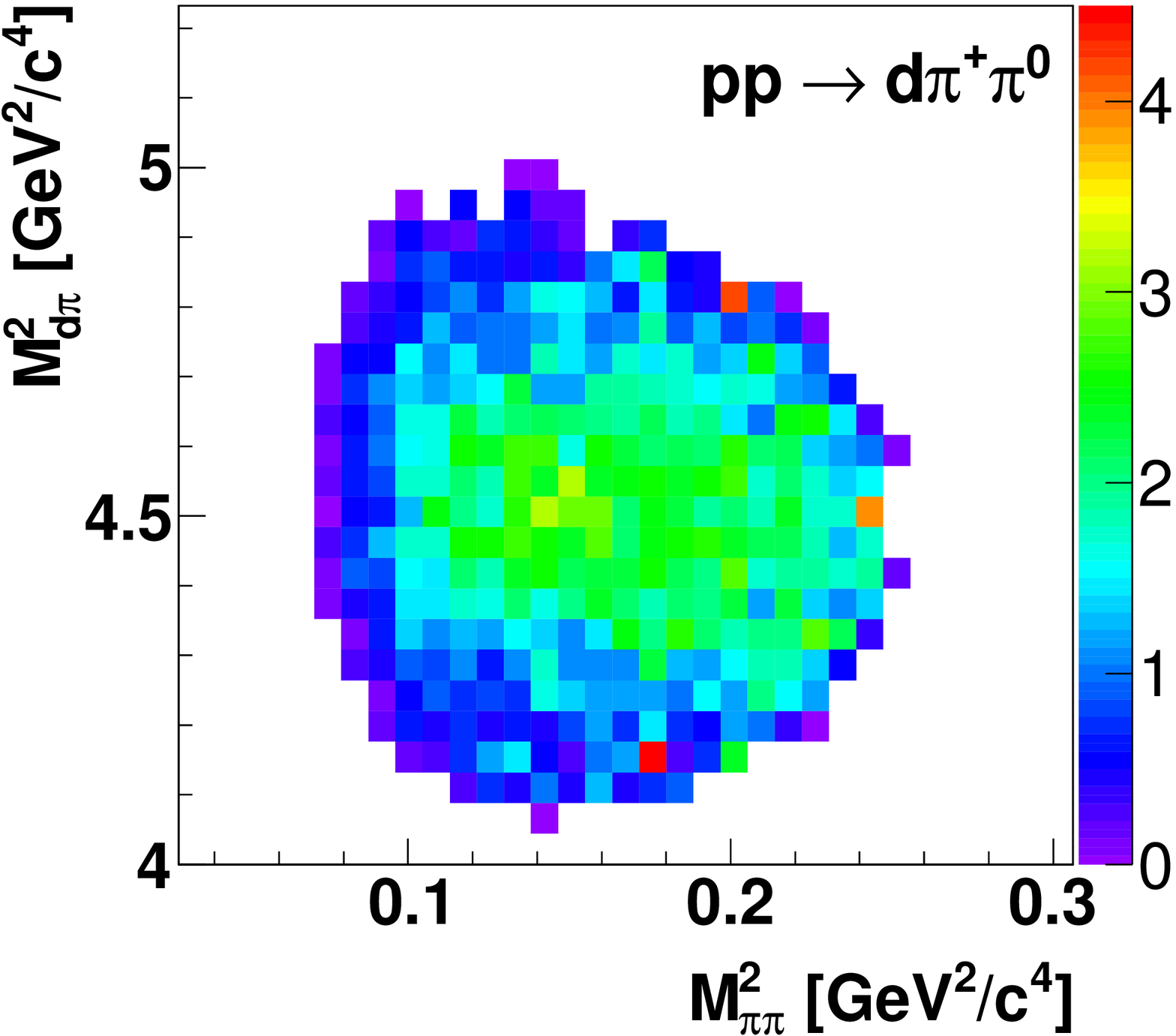}
\includegraphics[width=0.42\textwidth]{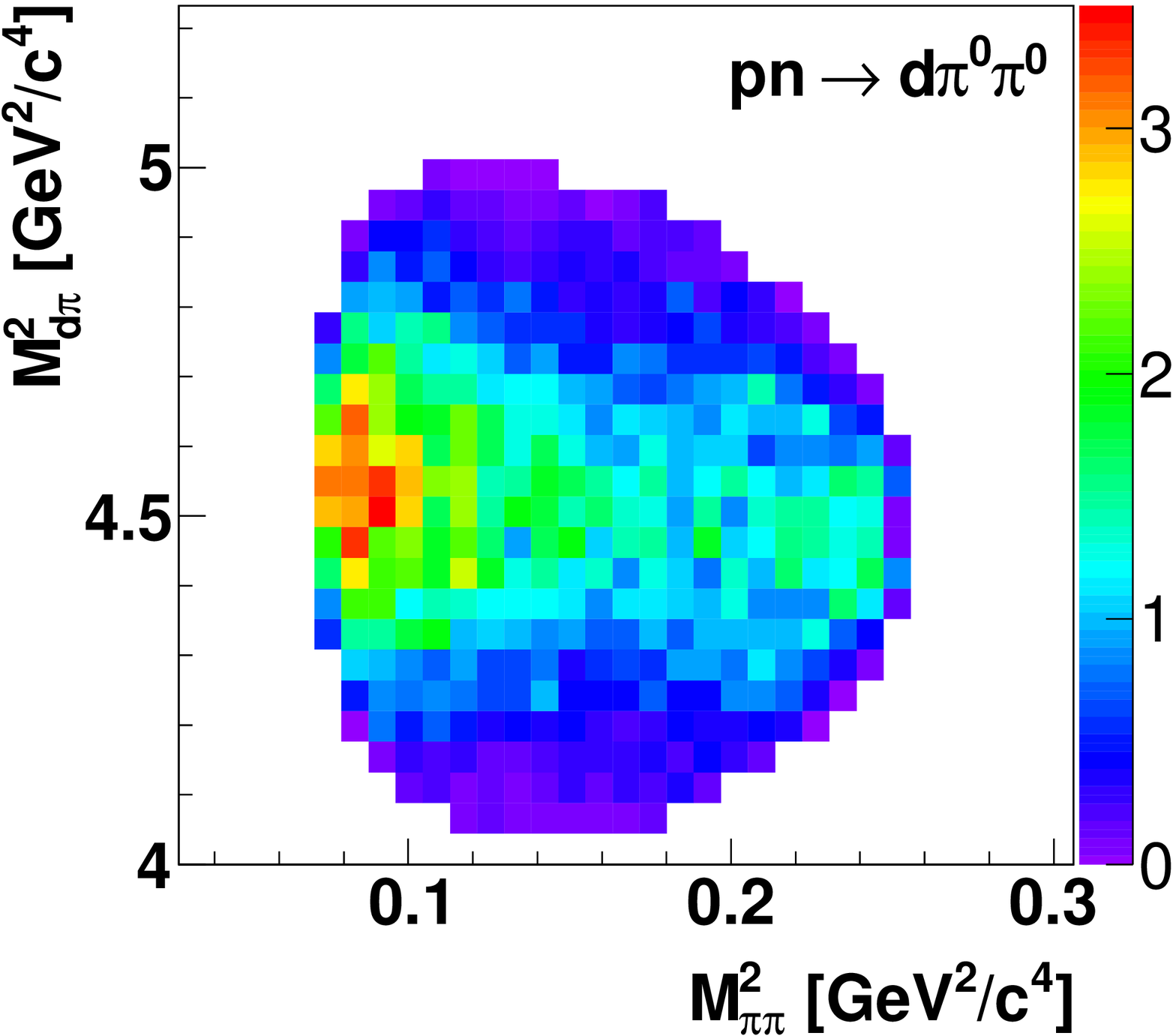}
\includegraphics[width=0.42\textwidth]{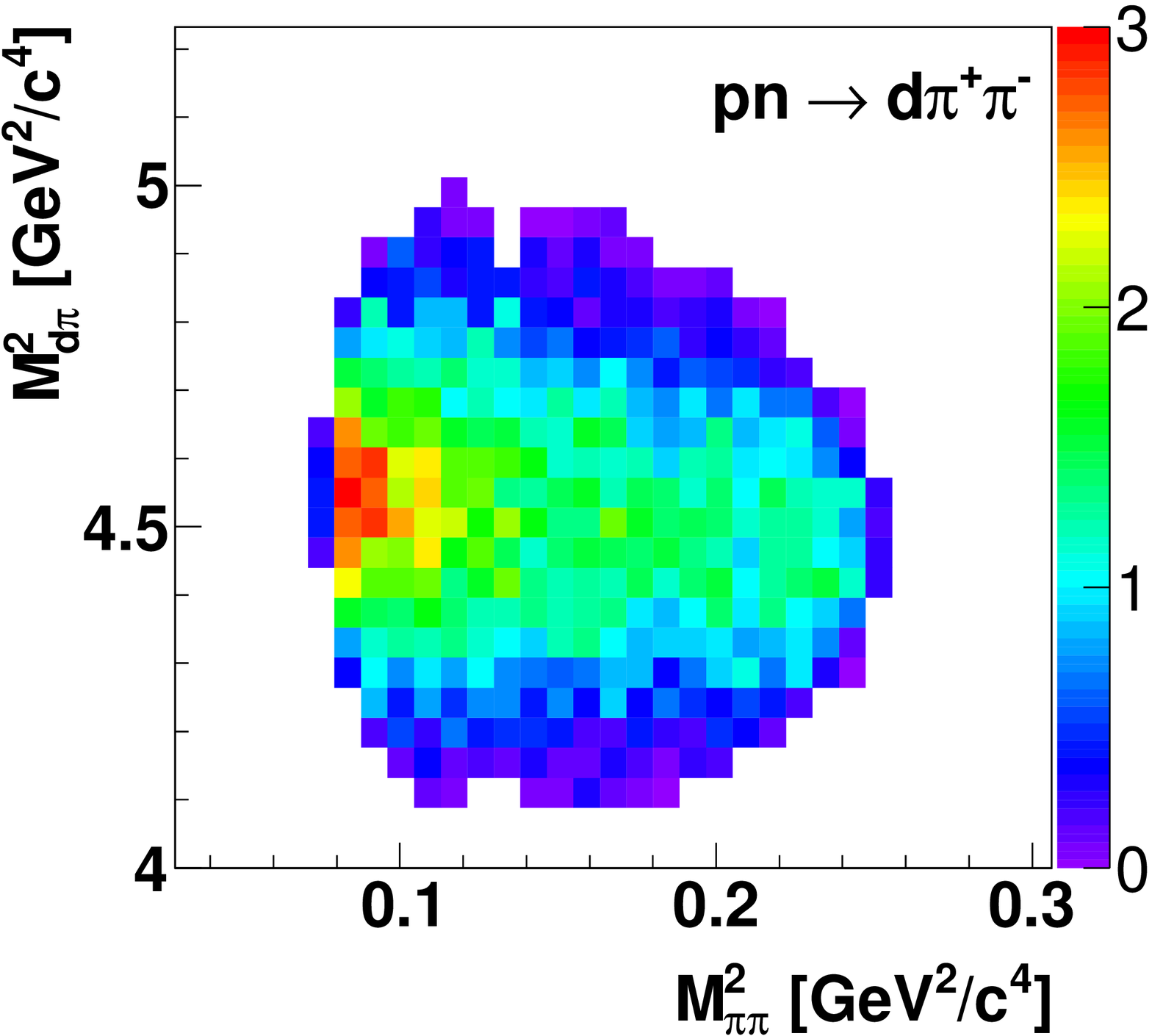}

\caption{
  Dalitz plots of $M_{d\pi}^2$ versus $M_{\pi\pi}^2$ at $\sqrt s$ = 2.37 GeV
  for the reactions $pp \to d\pi^+\pi^0$ (top, isovector), $pn \to
  d\pi^0\pi^0$  (middle, isoscalar) and  $pn~\to~d\pi^+\pi^-$ (bottom,
  isospin-mixed).  
}
\label{fig6}
\end{center}
\end{figure}


\begin{figure}[t]
\begin{center}

\includegraphics[width=0.235\textwidth]{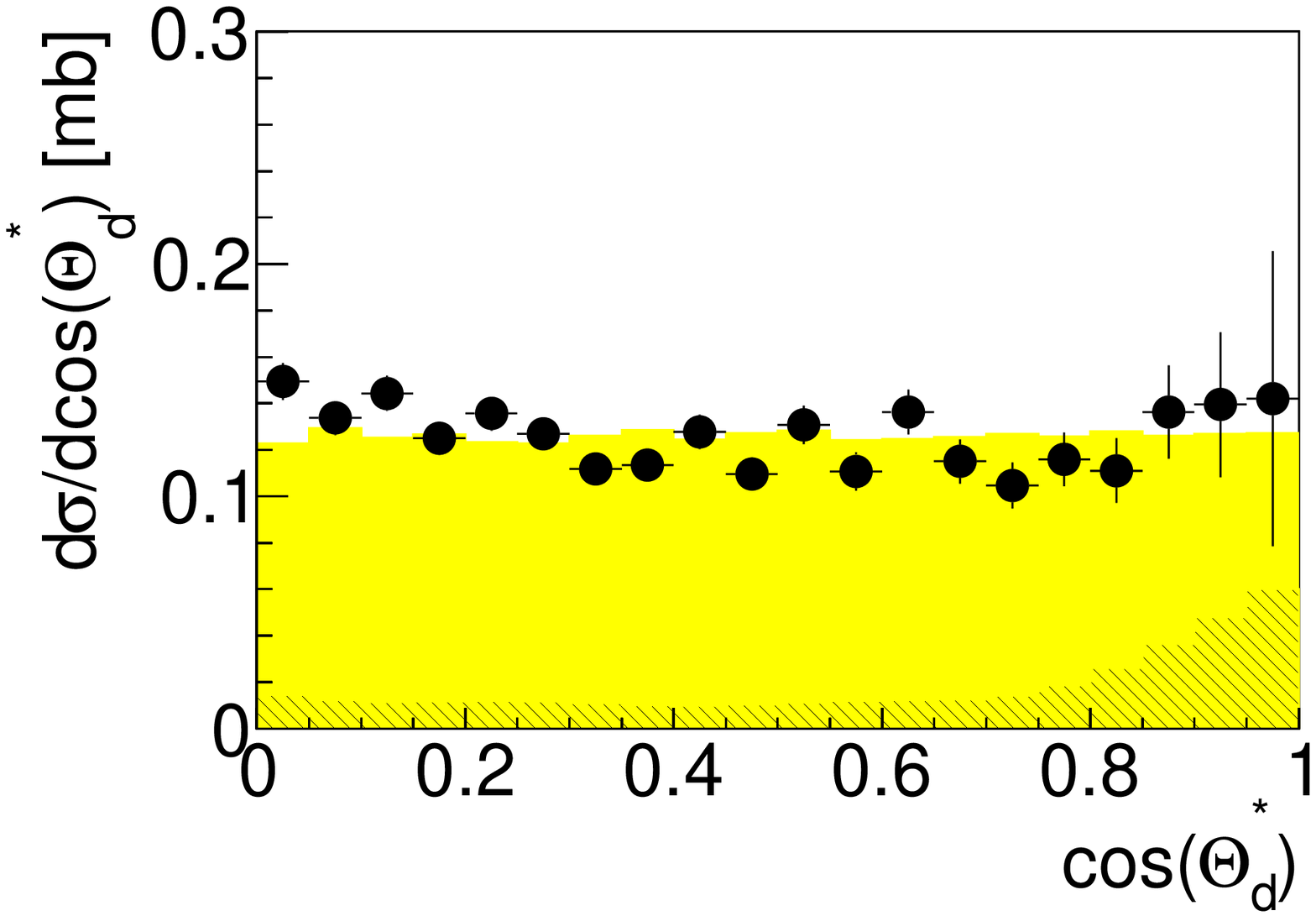}
\includegraphics[width=0.235\textwidth]{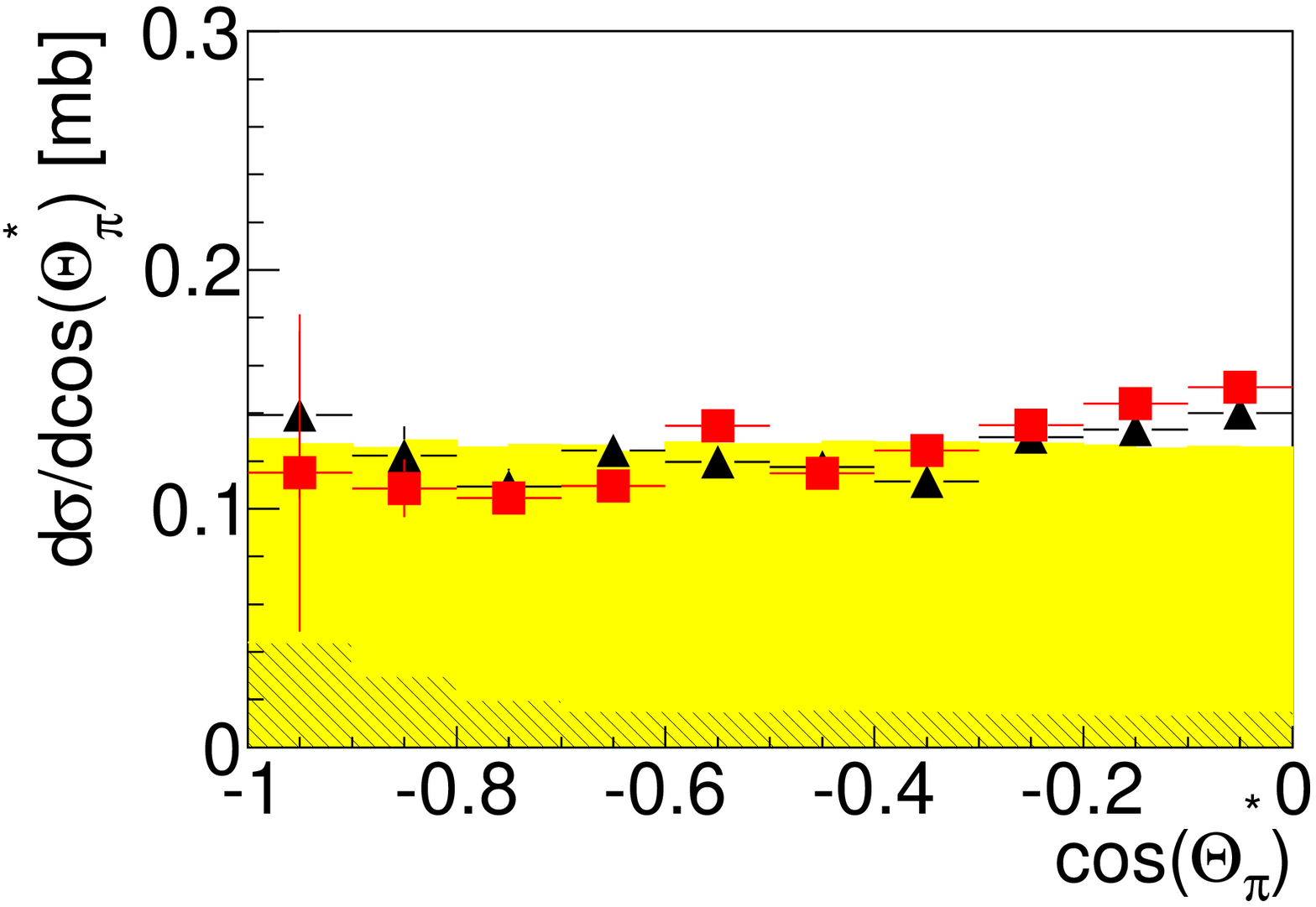}
\includegraphics[width=0.235\textwidth]{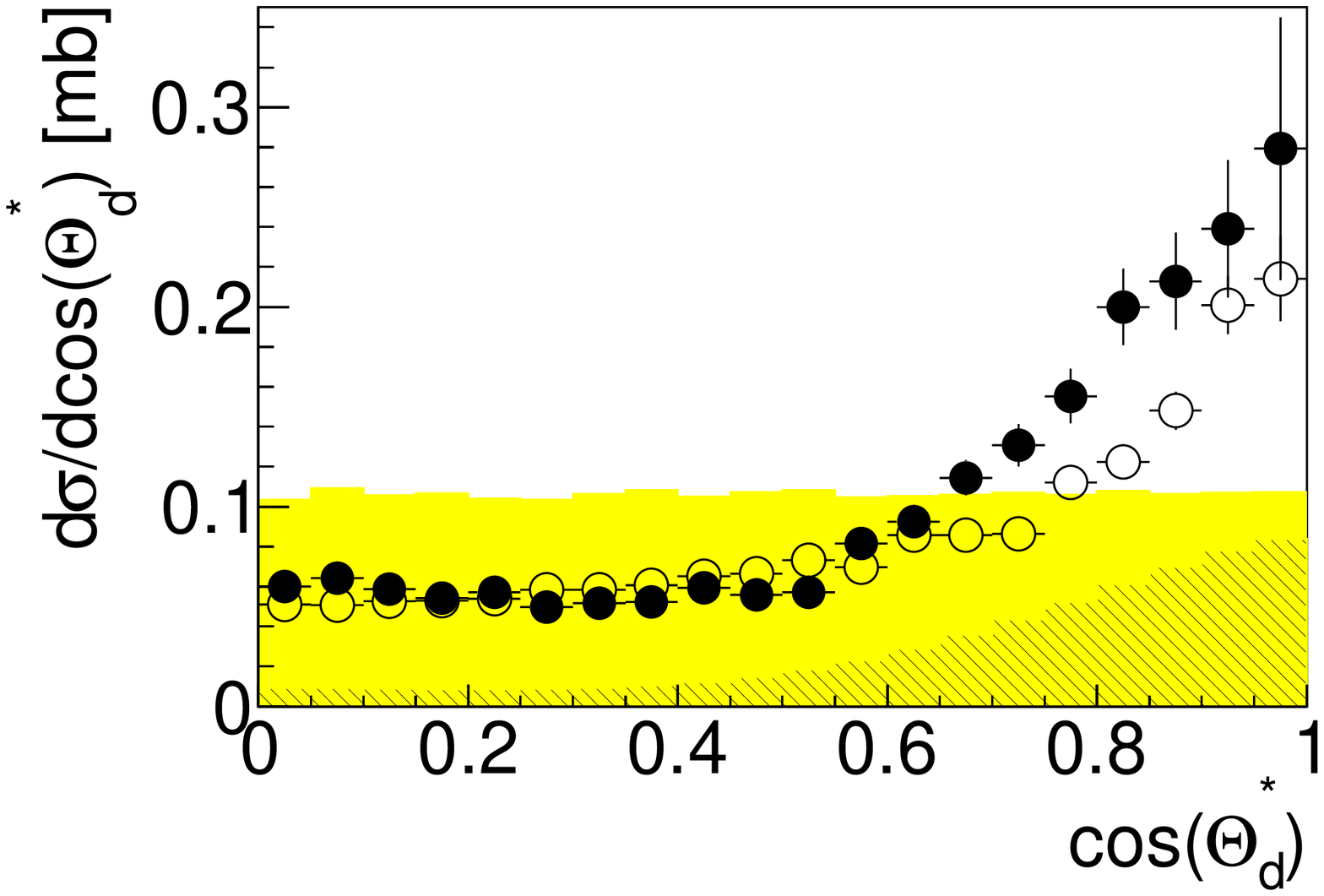}
\includegraphics[width=0.235\textwidth]{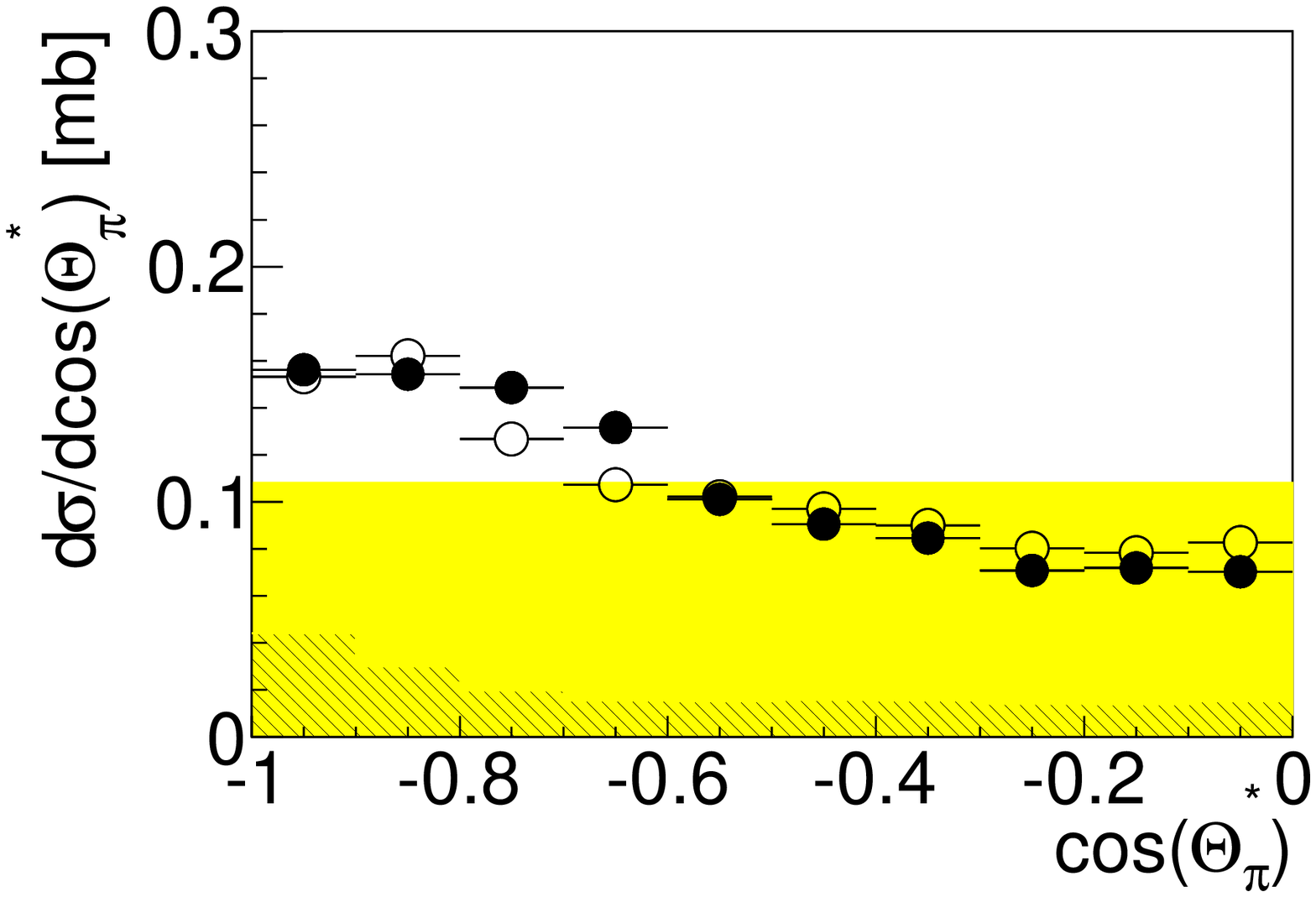}
\includegraphics[width=0.235\textwidth]{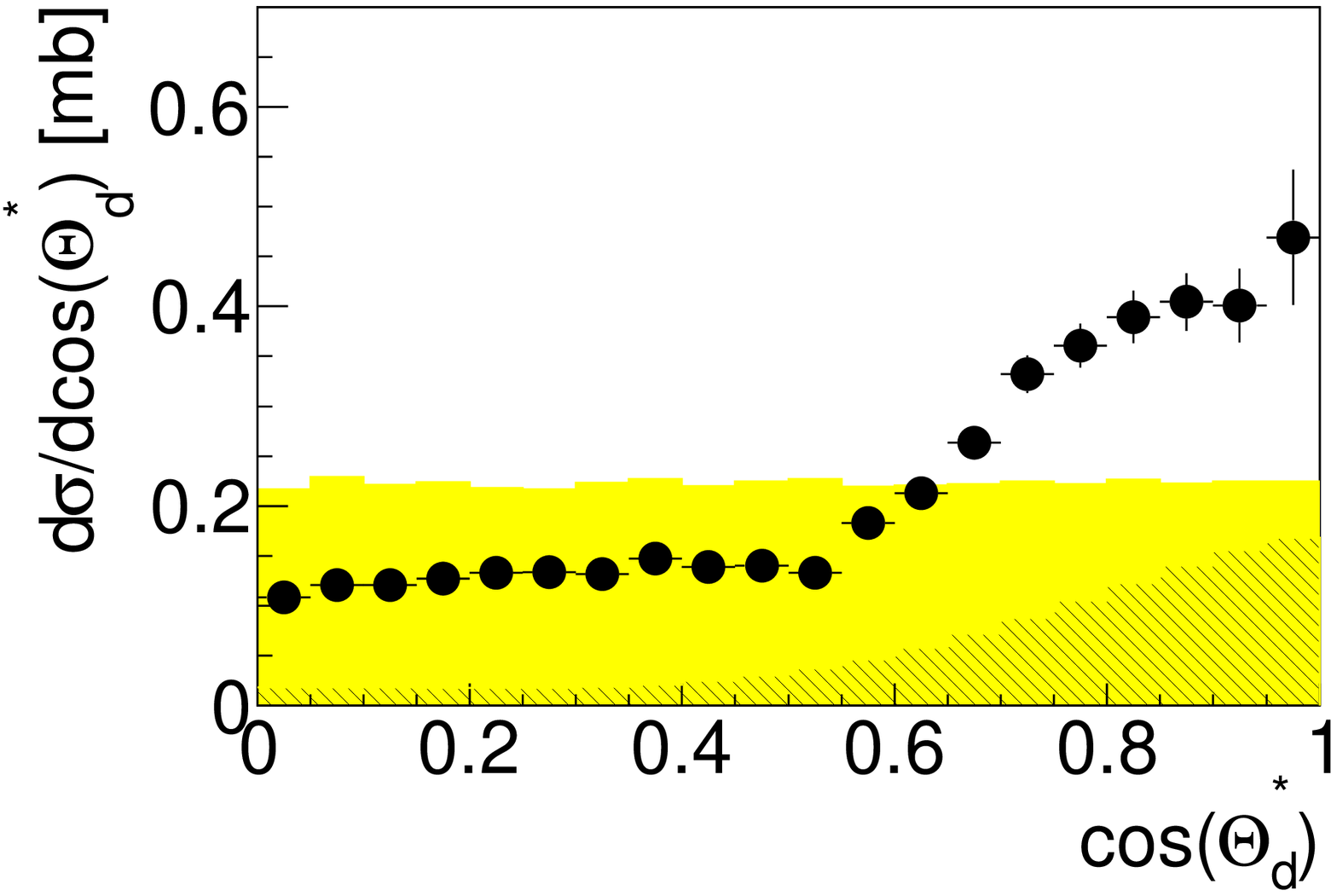}
\includegraphics[width=0.235\textwidth]{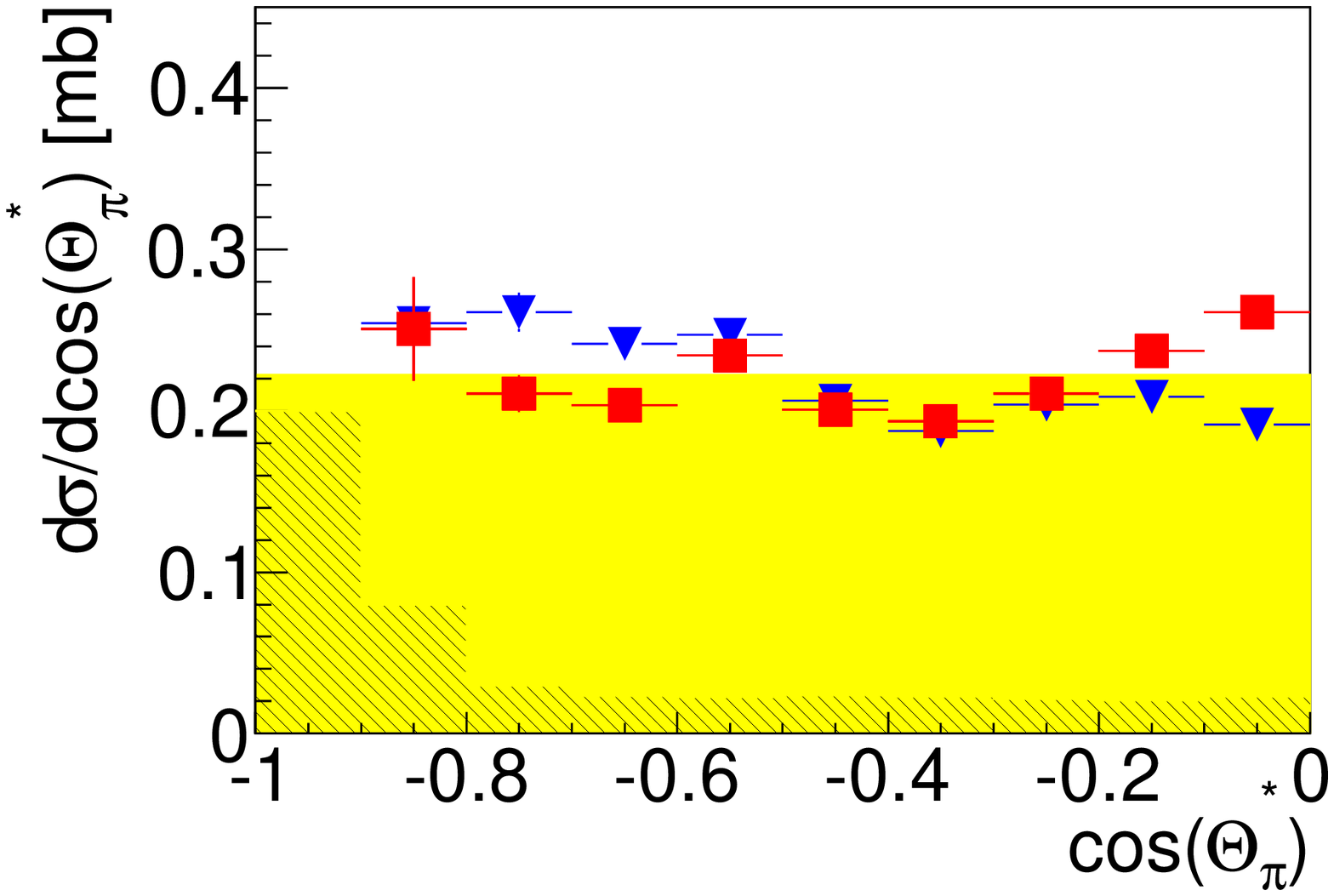}

\caption{
  Center-of-mass angular distributions for deuterons (left) and pions (right)
  at $\sqrt s$ = 2.37 GeV for reactions $pp \to d\pi^+\pi^0$ (top, isovector),
  $pn \to d\pi^0\pi^0$  (middle, isoscalar) and  $pn \to d\pi^+\pi^-$ (bottom,
  isospin-mixed). The filled circles denote the results from this
  work, the open circles the renormalized results from
  Ref. \cite{prl2011}. Filled squares denote $\pi^+$ distributions, filled
  triangles $\pi^0$ (top figure) and filled reversed triangles $\pi^-$ (bottom
  figure) distributions. Shaded and hatched areas represent phase-space
  distributions and systematic uncertainties, respectively.
}
\label{fig7}
\end{center}
\end{figure}

For all three channels the invariant-mass distributions are markedly different
from pure phase-space distributions, which are given in Figs.~3 and 4 by the
shaded histograms. 
As demonstrated by the six selected energy bins in Figs.~3 and 4, the
$M_{\pi^+\pi^-}$ spectra undergo quite some change in their shape across the
inspected energy region, although the $M_{\pi^0\pi^0}$ and $M_{\pi^+\pi^0}$
spectra  are stable in their shape over the considered region. The reason for
this can be understood from the fact that the isovector part grows
continuously and rapidly with increasing energy, thus increasing steadily its
influence in the $M_{\pi^+\pi^-}$  spectra. In consequence also the Dalitz
plot for the $pn \to d\pi^+\pi^-$  reaction changes accordingly over the energy region
considered -- as demonstrated for three energies in Fig.~5. 

In Fig.~6 the
Dalitz plots for all three reactions are compared at $\sqrt s$ = 2.37 GeV, the
peak cross section of the ABC effect. In all three cases we clearly observe
the horizontal band, which is due to the excitation of the $\Delta\Delta$
system. 
Hence opposite to the $M_{\pi\pi}^2$ distribution the $M_{d\pi}^2$
distribution is similar in all three cases.  

Finally we compare in Fig.~7 the center-of-mass (cms) angular distributions of
deuterons and pions. Whereas they are close to isotropic in the isovector
case -- as already noted in Ref. \cite{FK}, they are strongly anisotropic
in the isoscalar case. The latter was used 
for the determination of the spin $J$ = 3 of the isoscalar resonance structure
\cite{prl2011}.  The angular distributions for the $pn \to d\pi^+\pi^-$
reaction are in between both cases -- as expected for the isospin mixed
situation. Due to the reduced detection efficiency for charged particles at
small lab angles the systematic uncertainties (hatched histograms in Fig.~7)
are largest there. 

In the $pp \to d\pi^+\pi^0$ and $pn \to d\pi^+\pi^-$ reaction the charge states
of the produced pion pair are non-identical and hence can be
distinguished experimentally. We therefore plot in Fig.~7 the angular
distributions for both pions of the produced pair. As a result we see
that $\pi^+$ (filled squares) and $\pi^0$ (filled triangles) distributions
agree within uncertainties in case of the $pp \to d\pi^+\pi^0$ reaction. A 
similar situation is observed for the $\pi^+$ and $\pi^-$ (filled reversed
triangles) distributions in case of the $pn \to d\pi^+\pi^-$ reaction. This
suggests that the pion production mechanism should be identical for both 
pions of the produced pair -- as it is the case for a two-pion decay of an
exited nucleon state or an intermediate $\Delta\Delta$
system, which has been already postulated on the basis of the Dalitz plots.

\section{Summary and Conclusions}

The first exclusive and kinematically complete measurements of solid
statistics have been carried out for all three basic double-pionic fusion
channels simultaneously in the energy range  2.3 GeV $< \sqrt s <$ 2.5
GeV --- the energy region of the ABC effect and its associated narrow
resonance structure  around 2.37 GeV. This effort has been achieved by
exploiting quasifree kinematics with a proton beam impinging on a deuterium
target.  

The data for the isoscalar $d\pi^0\pi^0$ channel are in good agreement with
our previous measurement, though we find a substantially lower total cross
section for this channel. 
However, the 40$\%$ lower total cross 
section obtained now is still within the systematic uncertainties involved in
the absolute normalization of our data. 
Again, the data are characterized by the pronounced ABC effect associated with a
pronounced narrow resonance structure in the total cross section.

In contrast to the isoscalar $d\pi^0\pi^0$ channel the
isovector $d\pi^+\pi^0$ channel exhibits no ABC effect and no resonance
structure --- just a monotonically rising total cross section. 

The data for the isospin mixed $d\pi^+\pi^-$ channel agree as expected with
the combined results for isoscalar and isovector channels -- after accounting
for the isospin violation due to the different masses of neutral
and charged pions. The only previous data in this channel have been
low-statistics bubble-chamber measurements mainly using neutron beams of poor
momentum resolution. From inspection of Fig.~1 it gets immediately clear that
experiments under such conditions had no chance to discover the small narrow
structure in the total $pn \to d\pi^+\pi^-$ cross section, which gives the
hidden hint for the ABC resonance structure in this reaction. Hence it is not
surprising that it was left to the first measurements of the $\pi^0\pi^0$
production \cite{prl2011,MB} to reveal the 
resonance structure underlying the ABC effect in an environment of low
background from conventional processes.

\section{Acknowledgments}

We acknowledge valuable discussions with 
E. Oset,
A. Sibirtsev and C. Wilkin on this issue. 
This work has been supported by BMBF
(06TU9193), Forschungszentrum J\"ulich (COSY-FFE) and  
DFG (Europ. Graduiertenkolleg 683), the Swedish Research Council, the
Wallenberg foundation, the Foundation for Polish Science (MPD) and by the Polish
National Science Center
under grants No. 0320/B/H03/2011/40, 2011/01/B/ST2/00431,
2011/03/B/ST2/01847, 0312/B/H03/2011/40.
We also acknowledge the support from 
the EC-Research Infrastructure Integrating Activity `Study of Strongly
Interacting Matter' (HadronPhysics2, Grant Agreement n. 227431) under the
Seventh Framework Programme of EU. . 

}


\begin{thebibliography}{9}
\bibitem{abc} N. E. Booth, A. Abashian, K. M. Crowe, Phys. Rev. Lett.
  {\bf 7} (1961) 35; {\bf 6} (1960) 258; Phys. Rev. {\bf 132} (1963)
  2296ff.
\bibitem{prl2011} P. Adlarson et al., Phys. Rev. Lett. \textbf{106},
  (2011) 242302; arXiv: 1104.0123 [nucl-ex].
\bibitem{MB} M. Bashkanov et al.,  Phys. Rev. Lett. {\bf 102} (2009) 052301;
  arXiv: 0806.4942 [nucl-ex]. 
\bibitem{ris} T. Risser and M. D. Shuster, Phys. Lett. {\bf 43B}, (1973) 68.
\bibitem{FK} F. Kren et al., Phys. Lett. \textbf{B 684}, (2010) 110 and
  Erratum -- ibid. \textbf{B 702} (2011) 312; arXiv: 0910.0995v2 [nucl-ex].
\bibitem{shim} F. Shimizu et al., Nucl. Phys. {\bf A 386} (1982) 571.
\bibitem{bys} J. Bystricky et al., J. Physique {\bf 48} (1987) 1901.
\bibitem{alv} L. Alvarez-Ruso, E. Oset, E. Hernandez, Nucl. Phys. {\bf
   A633} (1998) 519 and priv. comm.; arXiv: 9706046 [nucl-th].
\bibitem{iso} T. Skorodko et al., Phys. Lett. {\bf B679} (2009) 30; arXiv:
  0906.3087 [nucl-ex].
\bibitem{deldel} T. Skorodko et al., Phys. Lett. \textbf{B 695}, (2011) 115;
  arXiv: 1007.0405 [nucl-ex]. 
\bibitem{nnpipi} T. Skorodko et al., Eur. Phys. J. \textbf{A 47}, (2011) 108;
  arXiv: 1012.1463 [nucl-ex]. 
\bibitem{bar} I. Bar-Nir et al., Nucl. Phys. {\bf B 54} (1973) 17.
\bibitem{abd} A. Abdivaliev  et al., Sov. J. Nucl. Phys. {\bf 29} (1979) 796.
\bibitem{plo} F. Plouin et al., Nucl. Phys. {\bf A 302} (1978) 413.
\bibitem{barg} Ch. Bargholtz et al., Nucl. Instrum. Methods
  \textbf{A 594} (2008) 339.
\bibitem{wasa} H.~H.~Adam et al., arXiv: 0411.038 [nucl-ex].
\bibitem{calen} H. Cal\'{e}n et al., Phys. Rev. {\bf C 58} (1998) 2667.



\end{thebibliography}
\end{document}